\documentstyle[psfig,graphicx]{mn}                       

\title[The near-infrared Ca\,{\sc ii} triplet -- III. Fitting functions]
{Empirical calibration of the near-infrared Ca\,{\large \bf II}
triplet -- III. Fitting functions}

\author[A.J. Cenarro et al.]  
    {A.J.~Cenarro,$^1$\thanks{E-mail: cen@astrax.fis.ucm.es}
    J.~Gorgas,$^1$ N.~Cardiel,$^1$ A.~Vazdekis,$^2$ and
    R.F.~Peletier$^3$\\ $^1$Depto. de Astrof\'{\i}sica, Fac. de
    Ciencias F\'{\i}sicas, Universidad Complutense de Madrid, E-28040
    Madrid, Spain.\\$^2$Instituto de Astrof\'{\i}sica de Canarias,
    E-38200, La Laguna, Tenerife, Spain.\\$^3$School of Physics and
    Astronomy, University of Nottingham, University Park, Nottingham
    NG7 2RD, UK\\}
\date{Accepted 2001 October 11. Received 2001 August 06}
\pagerange{\pageref{firstpage}--\pageref{lastpage}}
\pubyear{2000}

\def\LaTeX{L\kern-.36em\raise.3ex\hbox{a}\kern-.15em
    T\kern-.1667em\lower.7ex\hbox{E}\kern-.125emX}

\begin{document}

\label{firstpage}

\maketitle

\begin{abstract}

Using a near-IR stellar library of 706 stars with a wide coverage of
atmospheric parameters, we study the behaviour of the Ca\,{\sc ii}
triplet strength in terms of effective temperature, surface gravity
and metallicity. Empirical fitting functions for recently defined
line-strength indices, namely CaT$^*$, CaT and PaT, are
provided. These functions can be easily implemented into stellar
populations models to provide accurate predictions for integrated
Ca\,{\sc ii} strengths. We also present a thorough study of the
various error sources and their relation to the residuals of the
derived fitting functions.  Finally, the derived functional forms and
the behaviour of the predicted Ca\,{\sc ii} are compared with those of
previous works in the field.

\end{abstract}


\section{Introduction}

This is the third paper in a series dedicated to the understanding of
stellar populations of early--type galaxies and other stellar systems
by using their near--IR spectra and, particularly, the strength of the
integrated Ca\,{\sc ii} triplet. In the previous papers, we have
presented the basic ingredients of this project. First, we have
observed a new stellar library of 706 stars at 1.5~\AA\ (FWHM)
spectral resolution in the range $\lambda\lambda$~8348-9020~\AA\
(Cenarro et al. 2001a, Paper~I). That paper also includes the
definition of 3 new, improved line-strength indices in the Ca\,{\sc
ii} triplet region (namely CaT, PaT, CaT$^{*}$), which are especially
suited to be measured in the integrated spectra of stellar
populations. Paper~II of the series (Cenarro et al. 2001b) presents an
updated set of atmospheric parameters ($T_{{\rm eff}}$, $\log g$ and
[Fe/H]) for the stars of the library. The objective of this third
paper is to provide empirical fitting functions describing the
behaviour of the above indices in terms of the atmospheric
parameters. Together with the spectra of the stellar library, the
fitting functions will be implemented into an evolutionary stellar
populations synthesis code to predict both the spectral energy
distribution and the strengths of the indices for stellar populations
of several ages and metallicities (Vazdekis et al. 2001, Paper~IV).

A popular method to investigate the star formation history and the
stellar content of galaxies is to compare observed line strength
indices with stellar population models. This requires an accurate,
prior knowledge of the behaviour of the spectral features of interest
for a wide range of stellar spectral types, luminosity classes and
metallicities, which can be accomplished with the aid of either
empirical stellar libraries or theoretical model atmospheres. There
are two common alternatives for the stellar populations synthesis
approach. The first method mixes the spectra of the different stars
with their relative ratios given by evolutionary synthesis models to
predict spectral energy distributions (Fioc \& Rocca-Volmerange 1997;
Leitherer et al. 1999; Vazdekis 1999; Schiavon, Barbuy \& Bruzual
2000; Bruzual \& Charlot 2001). The major advantage of this spectral
synthesis is that it provides full information of all the spectral
features within the spectral range covered by the stellar
library. However, its usefulness relies on the availability of a
complete stellar library at the appropriate spectral resolution. The
second procedure, and probably the most widely employed in the past,
is the use of empirical fitting functions describing the strength of
previously defined spectral features in terms of the main atmospheric
parameters. These calibrations are directly implemented into the
stellar populations models to derive the index values for populations
of different ages and metallicities. It must be noted that one of the
main advantages of using fitting functions to reproduce the behaviour
of spectral indices is that it allows stellar populations models to
include the contribution of all the required stars by means of smooth
interpolations between well-populated regions in the parameter
space. One should not forget that, in any of the above two methods, a
common limitation of the empirical procedures arises from the fact
that they implicitly include the chemical enrichment history of the
solar neighborhood. This caveat must be kept in mind when using model
predictions to interpret the stellar populations of external galaxies,
whose star formation histories might be completely different to that
of the Galaxy.

The usefulness of the fitting functions approach has been clearly
demonstrated by the fact that the most important evolutionary
synthesis models (e.g., Worthey 1994, Vazdekis et al. 1996, Tantalo et
al. 1996 or Bruzual \& Charlot 2001) have implemented the available
fitting functions to fit observed line strengths in the
literature. Unfortunately, at present fitting functions are only
available in the blue and visible part of the spectrum (e.g. Gorgas et
al. 1993, Worthey et al. 1994 and Worthey \& Ottaviani 1997 for the
Lick/IDS indices; Poggianti \& Barbaro 1997 and Gorgas et al. 1999 for
the $\lambda$ 4000\AA\ break). Since other spectral regions also
contain very useful, complementary, absorption lines, it is necessary
to extend this kind of calibrations to line-strength indices in other
spectral regimes, such as the ultraviolet and the near infrared
(e.g. the CO index at 2.2~$\mu$m analyzed by Doyon, Joseph \& Wright
1994). This is indeed an important motivation for this paper. So,
before we give a comprehensive analysis of the predictions of both the
spectral synthesis and the fitting functions for stellar populations
(Paper IV), we will concentrate in this paper on understanding the
behaviour of the Ca\,{\sc ii} triplet as a function of the atmospheric
parameters, providing the corresponding fitting functions.

The Ca\,{\sc ii} triplet is one of the most prominent features in the
near-IR region of the spectrum of cool stars. Even though there are
many previous works dealing with the near-IR Ca\,{\sc ii} triplet (we
refer the reader to the short review presented in Section~2 of
Paper~I), the papers by D\'{\i}az, Terlevich \& Terlevich (1989), Zhou
(1991), J$\o$rgensen, Carlsson \& Johnson (1992) and Idiart, Thevenin
\& de Freitas-Pacheco (1997) (hereafter DTT, ZHO, JCJ and ITD,
respectively) deserve a special mention, since they provide empirical
fitting functions for the Ca\,{\sc ii} triplet. They have been very
useful to obtain a first understanding of the behaviour of the
Ca\,{\sc ii} triplet as a function of the stellar parameters but, as
we will see in this paper, suffer from several limitations that make
it impossible for stellar populations models to predict reliable
calcium strengths, especially for old stellar populations. Previous
papers which have made use of the above mentioned fitting functions to
predict the Ca\,{\sc ii} triplet in the integrated spectra of galaxies
include Vazdekis et al. (1996), ITD, Mayya (1997),
Garc\'{\i}a--Vargas, Moll\'{a} \& Bressan (1998), Leitherer et
al. (1999) and Moll\'{a} \& Garc\'{\i}a--Vargas (2000).

Section~2 of this paper describes the qualitative behaviour of the
Ca\,{\sc ii} triplet as a function of the atmospheric parameters, as
derived from the new stellar library. We also include a comparison
with the results of the previous work. We devote Section~3 to the
mathematical fitting procedure, providing the significant terms,
coefficients and statistics of the derived fitting
functions. Afterwards, a thorough analysis of residuals and possible
error sources is presented, including the sensitivity of the fitting
functions to differences in Ca/Fe ratios.  In Section~4, we compare
the new fitting functions with those presented in previous papers.
Finally, Sections~5 is reserved to discuss some important issues and
to summarize the contents of this paper.

\section{Behaviour of the Ca\,{\sc ii} triplet as a function of atmospheric parameters}
\label{stellib}

\begin{figure*}
\centerline{\hbox{
\psfig{figure=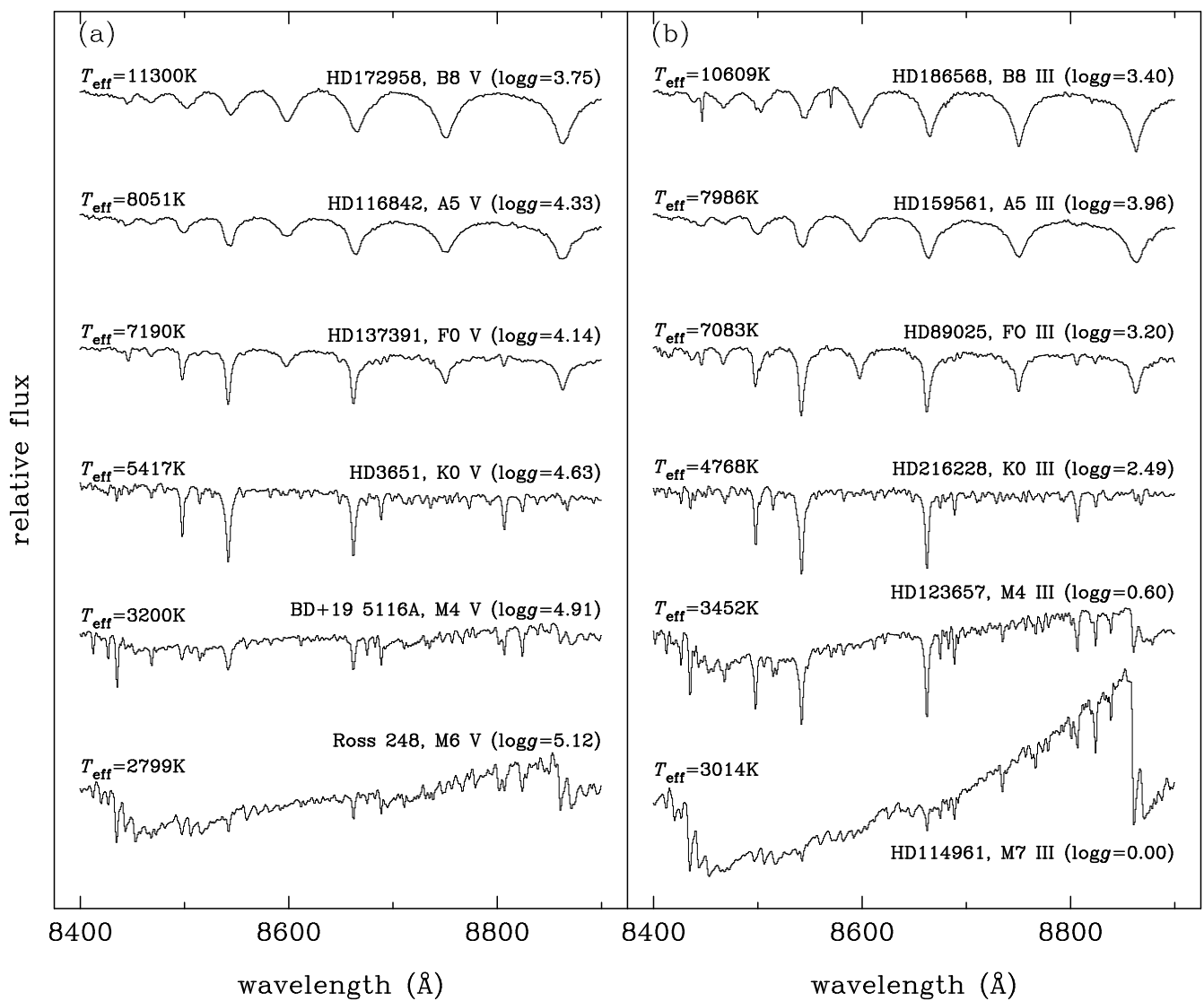}
}}   
\caption{Sequences in spectral types for (a) dwarf and (b) giant stars
from the near-IR stellar library. Effective temperatures, names,
spectral types, luminosity classes and surface gravities ($\log g$ in
dex) are given in the labels. All the spectra have been normalized and
reproduced using the same scales so relative differences among the
spectra are kept.}
\label{secTemp}
\end{figure*}

\begin{figure*}
\centerline{\hbox{
\psfig{figure=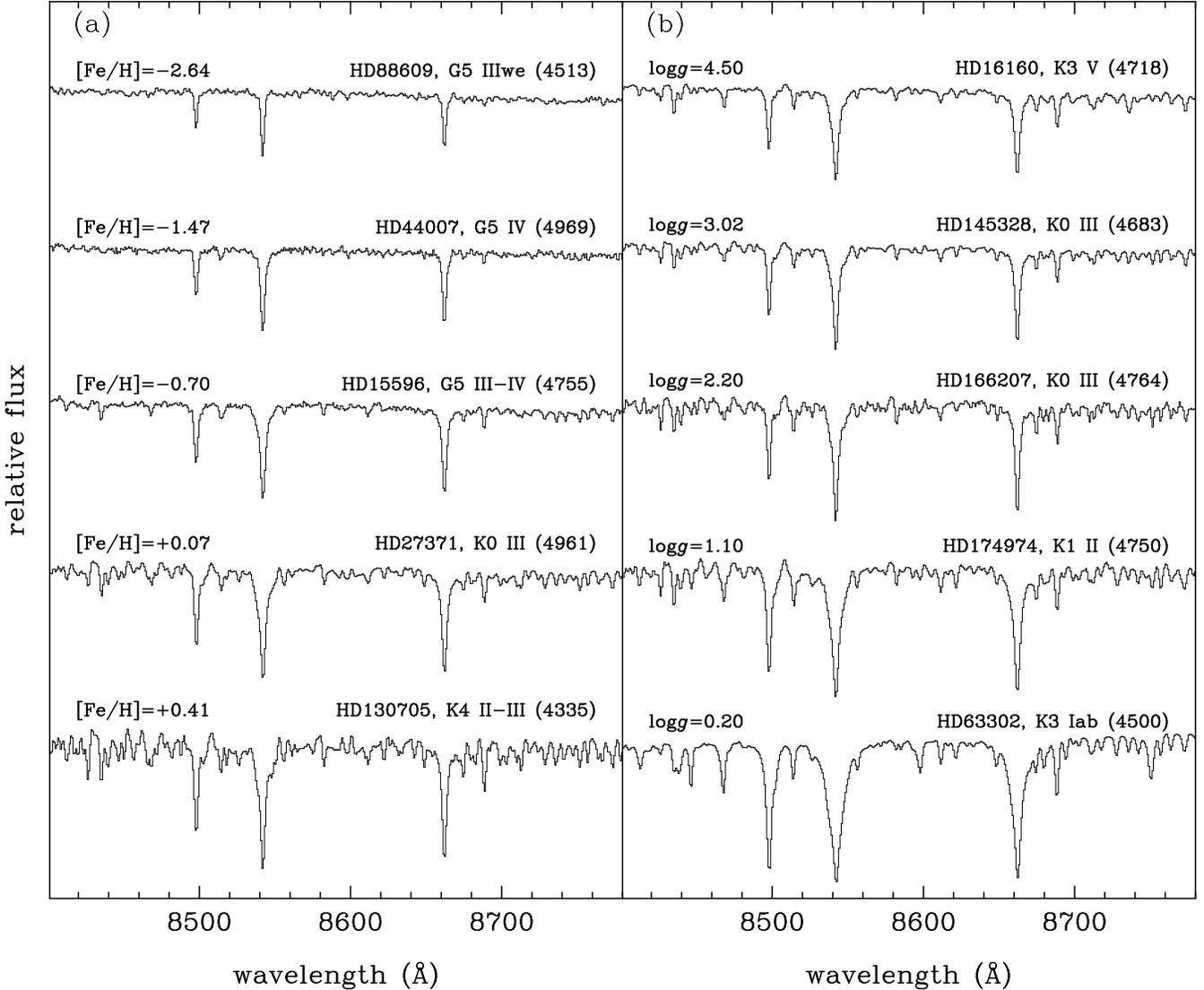}
}} 
\caption{Metallicity and gravity effects on the strength of the 
Ca\,{\sc ii} triplet for a subsample of stars from the near-IR stellar
library. Panel (a) shows stars with similar temperature and gravity
but spanning a wide range in metallicity. Panel (b) displays a
sequence in gravity for stars with similar temperature, and
metallicity around solar. Temperatures in K are given in brackets.
All the spectra have been normalized and reproduced using the same
scale.}
\label{secZlogg}
\end{figure*}

\begin{figure*}
\centerline{\hbox{
\psfig{figure=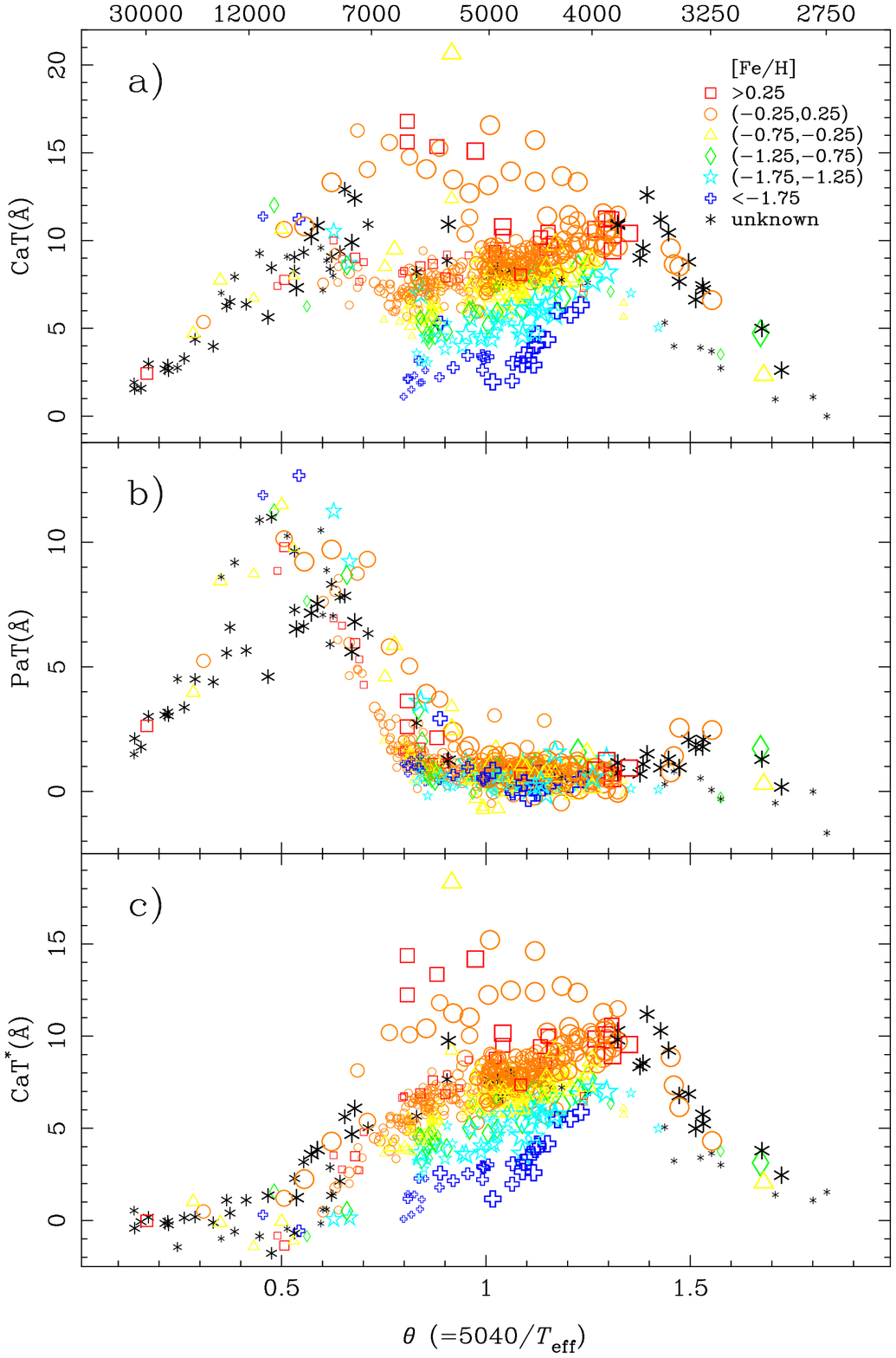}
}}
\caption{CaT, PaT and CaT$^{*} (= {\rm CaT}-0.93{\rm PaT})$ indices
versus $\theta (\equiv 5040/T_{\rm eff})$ for the whole stellar
library. Different symbols are used to indicate different
metallicities (as in the key), while sizes are related with surface
gravity, in the sense that the larger the symbol (supergiants), the
lower the gravity. On the top, the effective temperature scale is
given.}
\label{triplot}
\end{figure*}

\subsection{Qualitative behaviour}
\label{qualitative}

As a first step to understand the behaviour of the Ca\,{\sc ii}
triplet as a function of the atmospheric parameters, this section
describes, from a qualitative point of view, the effect of
temperature, surface gravity and metallicity on the strength of the
Ca\,{\sc ii} lines. Figure~\ref{secTemp} shows a comparative sequence
in spectral types for a sample of dwarfs (a) and giants (b) taken from
the new near-IR stellar library (see also Figure~1 in Paper~I for a
detailed description of the strongest spectral features). The effects
of metallicity and gravity on the Ca\,{\sc ii} lines are clearly shown
in Figure~\ref{secZlogg}. In Figure~\ref{triplot} we plot the measured
CaT, PaT and CaT$^{*}$ indices versus $\theta$ ($\equiv 5040/T_{\rm
eff}$) for the whole sample of stars.

\begin{table}
\centering{
\caption{Bandpass limits for the generic indices CaT and PaT.}
\label{defCaTPaT}
\begin{tabular}{@{}ccc@{}}
\hline                    
CaT central        & PaT central        & Continuum        \\                     
bandpasses (\AA)   & bandpasses (\AA)   & bandpasses (\AA) \\ 
\hline                     
Ca1 8484.0--8513.0 & Pa1 8461.0--8474.0 & 8474.0--8484.0   \\       
Ca2 8522.0--8562.0 & Pa2 8577.0--8619.0 & 8563.0--8577.0   \\       
Ca3 8642.0--8682.0 & Pa3 8730.0--8772.0 & 8619.0--8642.0   \\
                   &                    & 8700.0--8725.0   \\
                   &                    & 8776.0--8792.0   \\ 
\hline
\end{tabular}
}
\end{table}

It is worth reminding that the indices CaT and PaT measure,
respectively, the strengths of the raw calcium triplet and of three
pure H Paschen lines. They consist of five continuum bandpasses (which
are the same for both indices) and three central bandpasses covering,
in each case, the three Ca\,{\sc ii} lines (CaT) or the P17, P14 and
P12 lines of the Paschen series (PaT). Table~\ref{defCaTPaT} lists the
bandpass limits of these indices. The PaT index was specially
designed to quantify the CaT contamination with Paschen lines, and it
makes it possible to define the corrected index CaT$^{*} (= {\rm
CaT}-0.93{\rm PaT})$ as a reliable indicator of the pure Ca\,{\sc ii}
strength. We refer the reader to Paper~I for further details about the
definition and measurement of the above indices.

The spectra of hot stars (Fig.~\ref{secTemp}, top) are dominated by
Paschen lines, whose relative depths exhibit a smooth sequence with
wavelength, with the Ca\,{\sc ii} strengths, both in dwarfs and
giants, being clearly negligible (see also Fig.~\ref{triplot}c). As
temperature decreases, the Ca\,{\sc ii} lines become prominent,
standing out over the Paschen lines sequence (see the A and F types in
Fig.~\ref{secTemp}). The increasing strength of the Ca\,{\sc ii} index
with the decreasing temperature peaks at late K and early M types. For
a wide range of spectral types ($0.7 \leq \theta \leq 1.2$), the
Ca\,{\sc ii} lines are heavily affected by metallicity and gravity
effects (leading to the large spread of CaT and CaT$^{*}$ values in
Figs.~\ref{triplot}a and ~\ref{triplot}c), in the sense that their
strengths increase as metallicity increases and gravity decreases
(Fig.~\ref{secZlogg}). For lower temperatures, Ca goes gradually into
a neutral state and the Ca\,{\sc ii} strength decreases. The turning
point at which the trend with temperature changes depends on the
luminosity class, being cooler for giants ($\theta
\sim 1.3$) than for dwarfs ($\theta \sim 1.1$). Such an effect is
readily seen in Figs.~\ref{triplot}a and ~\ref{triplot}c, and also by
comparing the depths of the Ca\,{\sc ii} lines for the M4~V and M4~III
types in Fig.~\ref{secTemp}. Finally, strong molecular bands of TiO
and VO dominate the latest spectral types, in which the Ca\,{\sc ii}
strength is almost negligible (Fig.~\ref{secTemp}, bottom). For these
stars, as well as for the very hot ones, metallicity seems not to
affect the calcium strength.

The behaviour of the PaT index is illustrated in
Fig~\ref{triplot}b. As expected for a hydrogen index, it attains high
values for hot and warm stars ($\theta \la 0.8$), showing a maximum at
an effective temperature which depends on the luminosity class
(already reported by Andrillat, Jaschek \& Jaschek 1995). Even though,
in the low temperature regime, the stellar spectra do not present
Paschen lines, the index takes values slightly larger than zero. This
is mainly due to the contamination of weak metallic lines falling
within the central bands. Also, the presence of strong molecular bands
in the spectra of cold giants ($\theta \ga 1.4$) causes a fictitious
enlargement of the PaT values.

\subsection{Comparison with previous work}
\label{prevff}

In Paper~I we have already presented a brief review of the previous
work studying the Ca\,{\sc ii} triplet as a function of atmospheric
parameters. To facilitate the following discussion, in
Table~\ref{prevcal} we give a summary of the conclusions of these
previous studies. We include the main empirical papers on the subject,
together with the theoretical calibration of JCJ. Columns~3 to 5 give
the ranges of the atmospheric parameters spanned by the calibrating
stars (the numbers of stars used are listed in column~2) or atmosphere
models. The atmospheric parameters on which the Ca\,{\sc ii} triplet
was found to be dependent in these previous works are indicated in
columns 6 to 8, where the keywords {\sl strong\/} and {\sl weak\/} are
used to remark that the Ca\,{\sc ii} strength was reported to be more
sensitive to one parameter than to another.

\begin{table*}
\centering{
\caption{Previous calibrations of the Ca\,{\sc ii} triplet. Sources are the 
following: 
JAJ (Jones, Alloin \& Jones 1984), CVP (Carter, Visvanathan \& Pickles
1986), A\&B (Alloin \& Bica 1989), DTT (D\'{\i}az, Terlevich \& Terlevich
1989), ZHO (Zhou 1991), MAL (Mallik 1997), ITD (Idiart, Th\'{e}venin \& de
Freitas Pacheco 1997), and JCJ (J{\o}rgensen, Carlsson \& Johnson 1992).}
\label{prevcal}
\begin{tabular}{@{}lrcrrcccl@{}}
\hline
Source  &  \multicolumn{1}{c}{No.} & $\theta$ & \multicolumn{1}{c}{$\log g$} & 
\multicolumn{1}{c}{[Fe/H]} &
$T_{\rm eff}$ & $g$ & [Fe/H] & Fitting functions\\
 & \multicolumn{1}{c}{stars} & range & \multicolumn{1}{c}{range} & 
\multicolumn{1}{c}{range} & 
\multicolumn{3}{c}{dependences} & parameters \\\hline
JAJ  & 62  & $0.68 - 1.47$ & $0.7 - 4.8$  & $-0.60 - +0.43$ & no & strong & weak & 
\\
CVP  & 51  & $0.93 - 1.68$ & $0.0 - 4.6$  & $0 - +0.50$ & yes & yes & no &
\\
A\&B & 63  & $0.68 - 1.46$ & $0.7 - 4.8$  & $-0.60 - +0.43$ & ? & yes & yes &
\\
DTT  & 106 & $0.74 - 1.43$ & $0.2 - 4.6$  & $-2.70 - +0.55$ & no & yes & yes &
$\log g$, [Fe/H]\\
ZHO  & 144 & $0.76 - 1.72$ & $0.7 - 4.8$  & $-2.28 - +0.60$ & yes & yes & yes &
$(R-I)$, $(R-I)^2$, $\log g$, $\log^2 g$, $10^{\rm [Fe/H]}$\\
MAL  & 146 & $0.81 - 1.58$ & $-0.6 - 4.8$ & $ -3.0 - +1.01$ & weak & yes & yes &
\\
ITD  & 67  & $0.74 - 1.20$ & $0.8 - 4.5$  & $-3.15 - +0.35$ & yes & weak & strong &
$\theta$, $\log g$, [Fe/H], $\theta{\rm [Fe/H]}$, $\log g{\rm [Fe/H]}$, [Ca/Fe]\\
JCJ  & \multicolumn{1}{c}{--}  & $0.76 - 1.26$ & $0.0 - 4.0$      & $-1.0 - +0.20 $ & yes & yes & yes &
$T_{\rm eff}$, $T_{\rm eff}^2$, $\log g$, $\log^2 g$, $T_{\rm eff}\log g$ 
\\
\hline
\end{tabular}
}
\end{table*}

A first glance at Table~\ref{prevcal} reveals that there are several
apparent inconsistencies among the previous papers. These discordant
results are mainly due to differences in the coverage of the parameter
space and the index definitions. One of the points of discrepancy is
the sensitivity to effective temperature. For the low temperatures, as
it was first reported by Cohen (1978), Figs.~\ref{triplot}a and
\ref{triplot}c clearly show that the Ca\,{\sc ii} strength decreases
for stars colder than $\theta\approx1.3$. This behaviour has been thus
acknowledged by the papers which include a good number of M stars,
like Carter, Visvanathan \& Pickles (1986) and ZHO. However, other
authors, in particular Jones, Alloin \& Jones (1984; hereafter JAJ)
and DTT, were not able to detect such an effect. The problem here is
that the index definitions that they used are not well suited for
stars with strong TiO absorption bands (see both Section~4.2 and
Figure~2 of Paper~I). In the first case (JAJ), the index is uncertain
for cold stars since the red sideband falls in a strong TiO
absorption. In the second one (DTT), the position of the local
continuum causes the index to overestimate the Ca\,{\sc ii}
absorption. Concerning the hotter stars, Fig.~\ref{triplot}c shows
that the Ca\,{\sc ii} strengths decrease when $T_{\rm eff}$
increases. However, and with the exception of Alloin \& Bica (1989;
hereafter A\&B), this behaviour has not been noted in the previous
empirical papers since the Ca\,{\sc ii} lines are heavily contaminated
by the Paschen series for $\theta\leq0.8$ (compare
Figs.~\ref{triplot}a and
\ref{triplot}c). Note that JCJ do reproduce this dependence since they
compute synthetic calcium equivalent widths without the inclusion of
Paschen lines.

Concerning the gravity dependence, Table~\ref{prevcal} shows that most
previous papers have remarked the strong luminosity sensitivity of the
Ca\,{\sc ii} triplet. One exception is the work by ITD, in which they
found a weaker dependence on gravity. However, it should be noticed
that their sample comprises only a small number (8) of supergiants,
and that most of them (7) exhibit very low metallicities
($-2.6\leq{\rm [Fe/H]}\leq-1.4$). In other words, gravity and
metallicity effects may not be well separated. Note also that, as it
is shown below, gravity effects at the low metallicity range are
milder.  The discrepancies among different works, as far as the
sensitivity to metallicity is concerned, are clearly due to the [Fe/H]
range spanned by the calibrating stars. While the early works by JAJ
and Carter et al. (1986) could not detect a significant metallicity
dependence, subsequent papers which explored a broader [Fe/H] range
(like DTT and ITD) clearly showed that metal abundance is also a key
parameter.

\subsection{The gravity and metallicity dependences}

\begin{figure}
\centerline{\hbox{
\psfig{figure=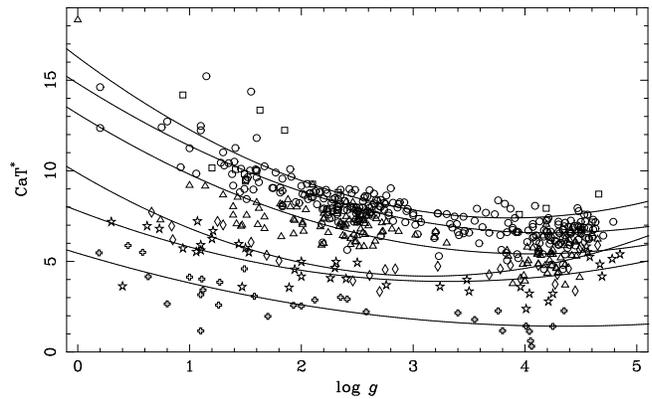}
}} 
\caption{Gravity dependence of the CaT$^{*}$ index for stars in the
temperature range $0.8\leq\theta\leq1.3$. Symbols types indicate
different metallicities following the code of Fig.~\ref{triplot}a. The
lines represent quadratic fits in $\log g$ for stars in the
metallicity ranges: ${\rm [Fe/H]}>0.25$, $-0.25<{\rm [Fe/H]}\leq0.25$,
$-0.75<{\rm [Fe/H]}\leq-0.25$, $-1.25<{\rm [Fe/H]}\leq-0.75$,
$-1.75<{\rm [Fe/H]}\leq-1.25$, and ${\rm [Fe/H]}\leq-1.75$ (from top
to bottom).}
\label{fig_g}
\end{figure}

In order to explore in more detail the interrelationship between
gravity and metallicity effects on the Ca\,{\sc ii} triplet of the
cool stars, we plot in Figure~\ref{fig_g} the gravity dependence for
different metallicity ranges, for stars with effective temperatures
$0.8\leq\theta\leq1.3$. To guide the eye, we have fitted parabola's
for each metallicity bin (note that these lines are not the definitive
fitting functions), checking that the second order term is
statistically significant in all cases. From this figure, it is clear
than the dependence on gravity is not linear. This shows that the
Ca\,{\sc ii} triplet is much more sensitive to gravity for giants and
supergiants than for dwarfs (for $\log g \ga 3.5$ the relation is
almost flat). This behaviour contrasts with the results of JAJ, A\&B
and DTT, who find a linear relation with $\log g$, but it is in
qualitative agreement with ZHO and, especially, with the conclusions
of Mallik (1997; hereafter MAL). Another result of Fig.~\ref{fig_g} is
that the gravity dependence tends to flatten off for low
metallicities. This effect was already noted by DTT when they
concluded that, at low $Z$, metallicity was the main parameter and it
is, again, in perfect agreement with the results of MAL (see their
Figure~6 and, also, Mallik 1994). It must also be noted that our
conclusions about the gravity dependence also agree extremely well
with the qualitative predictions of the theoretical study by JCJ. We
even notice their result about a Ca\,{\sc ii} increase at high
gravities for metallicities around ${\rm [Fe/H]}=-1$.

\begin{figure}
\centerline{\hbox{
\psfig{figure=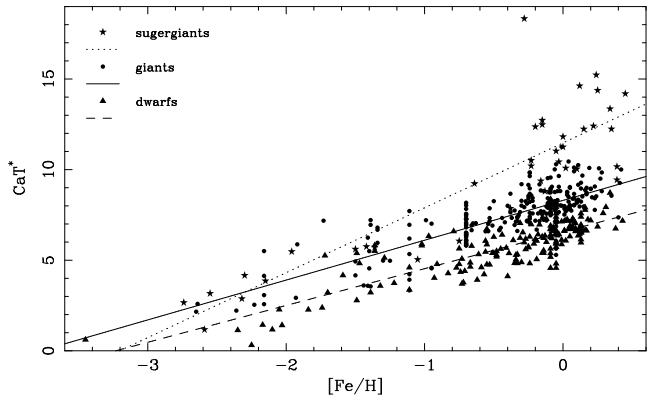}
}} 
\caption{Metallicity dependence of the CaT$^{*}$ index for stars in the
temperature range $0.8\leq\theta\leq1.3$. Different symbols are used
to indicate different luminosity classes, as shown in the key. The
straight lines are linear fits to each set of stars, with slopes of
$3.6\pm0.3$, $2.2\pm0.1$ and $2.0\pm0.1$, for supergiants, giants and
dwarfs, respectively}
\label{fig_z}
\end{figure}

The metallicity dependence is analysed in more detail in
Figure~\ref{fig_z}, where we have grouped our cool stars into three
luminosity classes. We also show linear fits to the [Fe/H] dependence
of each gravity subsample (in this case, the second order terms are
not statistically significant). The CaT$^*$--metallicity relation is
thus roughly linear for the cool stars, in agreement with A\&B.
Furthermore, we must note that we do not find any flattening of the
metallicity relation at high [Fe/H], as was reported by DTT (although
they derived a linear relation with [Fe/H], they concluded that, in
the high metallicity range, the Ca\,{\sc ii} strength is a function of
gravity only). More interestingly, we find that the slope of the
metallicity relation is statistically the same for giants and dwarfs
(already noticed by A\&B), but much steeper for the supergiant
subsample. Again, this is in agreement with the findings of MAL. In
particular, at low metallicities, giants and supergiants do not differ
in their Ca\,{\sc ii} strengths (as it was already apparent in
Fig~\ref{fig_g}).

To summarize, we find that the Ca\,{\sc ii} triplet in cool stars
($\sim$ F7 -- M0 spectral types) follows a complex dependence on
temperature, metallicity and gravity. Our conclusions are in very good
agreement with the empirical results of MAL and the theoretical
predictions of JCJ.

\section{The fitting functions}
\label{indexdef}

In this section we present the empirical fitting functions for the new
indices, as well as the general procedure followed to compute
them. These fitting functions have been calculated using the index
measurements given in Paper~I and the atmospheric parameters derived
in Paper~II. The complete tables of indices and parameters are
available at the URL addresses:\\ {\tt
http://www.ucm.es/info/Astrof/ellipt/CATRIPLET.html}\\ and\\ {\tt
http://www.nottingham.ac.uk/\~{}ppzrfp/CATRIPLET.html}

\subsection{The fitting procedure}

The main objective of this paper is to derive empirical fitting
functions for the indices CaT$^*$, CaT and PaT in terms of the stellar
atmospheric parameters. Keeping in mind that any of the above indices
can be expressed as a linear combination of the other two ones, we
have only computed fitting functions for CaT$^*$ and PaT. We chose
them since these indices are better measurements of pure Ca\,{\sc ii}
and H line-strengths respectively than CaT and, therefore, their
behaviours with atmospheric parameters are expected to be simpler than
in the case of a mixed line-strength like CaT. In any case, it is
important to clarify that the empirical calibrations are only
mathematical representations of the behaviour of the indices as a
function of the atmospheric parameters and, thus, a physical
justification of the derived coefficients is beyond the scope of this
paper.

Given the wide range of spectral types in the stellar library we
prefer to use $\theta$ as the effective temperature indicator,
together with $\log g$ and [Fe/H], the classical parameters for
surface gravity and metallicity. Following Gorgas et al. (1999) (see
also Gorgas et al. 1993, and Worthey et al. 1994), the fitting
functions have been computed as polynomials of the atmospheric
parameters in two possible functional forms,
\begin{equation}
\label{ffpol}
{\cal I} _{\rm a}(\theta, \log g, {\rm [Fe/H]}) = p(\theta, \log g, {\rm [Fe/H]}), \;\;\;\; {\rm or} 
\end{equation}
\begin{equation}
\label{ffexp}
{\cal I} _{\rm a}(\theta, \log g, {\rm [Fe/H]}) = {\rm const.} + \exp{\left[p(\theta, \log g, {\rm [Fe/H]})\right]} , 
\end{equation}
We chose the one that minimizes the residuals of the fit. ${\cal I}
_{\rm a}$ refers to any of the above indices and $p$ is a polynomial
with terms of up to the third order, including all possible
cross--terms among the parameters, i.e.,
\begin{equation}
p(\theta, \log g, {\rm [Fe/H]}) = \sum_{0 \le i+j+k \le 3}c_{i,j,k}\theta^{i}(\log g)^{j}{\rm [Fe/H]}^{k}, 
\label{eq3}
\end{equation}
with $0 \leq i+j+k \leq 3$  and $0 \leq i,j,k$.

Note however that, as a result of the wide parameter space covered by
the stellar library, there exists no single function of the forms
given by Eqs.~(\ref{ffpol}) and~(\ref{ffexp}) which can accurately
reproduce the complex behaviour of the indices CaT$^*$ and PaT. We
have therefore divided the whole parameter space into several ranges
of parameters (boxes) in which local fitting functions can be properly
computed. Finally, a general fitting function for the whole parameter
space has been constructed by interpolating the derived local
functions. To do so, we have defined the boundaries of the boxes in
such a way that they overlapped, including several stars in common. In
the overlapping zones cosine-weighted means of the functions
corresponding to both boxes were performed. This would mean that, if
${\cal I}_{\rm a 1}(x,y,z)$ and ${\cal I}_{\rm a 2}(x,y,z)$ are two
local fitting functions overlapping in the generic parameter $z$, and
defined respectively in the intervals $(z_{1,1},z_{1,2})$ and
$(z_{2,1},z_{2,2})$ with $z_{2,1}<z_{1,2}$, the predicted index in the
intermediate region will be given by
\begin{equation}
{\cal I} _{\rm a}(x,y,z) = w\,{\cal I} _{\rm a 1}(x,y,z) + (1-w)\,{\cal I} _{\rm a 2}(x,y,z),
\end{equation}
where the weight $w$ is modulated by the distance to the overlapping limits as
\begin{equation}
\label{weight}
w = \cos\left[\frac{\pi}{2}\left(\frac{z-z_{2,1}}{z_{1,2}-z_{2,1}}\right)\right],\,\,\,
{\rm with}\,\, z_{2,1}\leq z \leq z_{1,2}.
\end{equation}
After trying several functional forms for the weights, we confirmed
that the analytical expression given in Eq.~(\ref{weight}) guarantees,
in most cases, a smooth interpolation between local functions and
preserves the quality of the final fit.

The local fitting functions were derived through a weighted least
squares fit to all the stars within each parameter box, with weights
according to the uncertainties of the indices for each individual
star, as computed in Section~5 of Paper~I.  Since not all possible
terms of Eq.~(\ref{eq3}) were necessary, we followed a systematic
procedure to obtain the appropriate local fitting function in each
case.  To begin with, a general fit to all the 20 possible terms is
computed together with the residual variance of the fit and the
coefficients errors. The significance of each term is then calculated
by means of a $t$-test (that is, using the error in that coefficient
to check whether it is significantly different from zero) and the term
with the highest significance level ($\alpha$) is removed from the
fit. The whole procedure in then iterated, removing in each turn the
least significant coefficient, until all remaining coefficients are
statistically significant.  Typically, we have used a threshold value
of $\alpha = 0.10$ to keep a significant term. It must be noted that
the above procedure does not guarantee that we are obtaining the best
solution, so alternative approaches, like an inverse one (starting
from a one parameter fit and introducing the most significant term in
each iteration) were also followed to ensure that the final
combination of terms was the one which provided the minimum unbiased
residual variance.

\begin{table}                                                              
\centering{                                                                
\caption{Stars which were not used for the fitting functions
computation, coded as: C, carbon star; EmL, emission lines of the
elements in brackets; P, pulsating star; SB, spectroscopic binary;
Var, variable star; *: bad quality spectrum because of low
signal-to-noise ratios, bad exposing conditions, unreliable spectral
features, and others.}
\label{rejected}                                                            
\begin{tabular}{llll}          
\hline                                
Name & Diagnostic & Name & Diagnostic\\
\hline
HD 108   & EmL (Ca,H)           & HD 120933  & Var (CVn)   \\     
HD 1326B & Flare star           & HD 121447  & Var         \\
HD 17491 & P                    & HD 138279  & Var (RR Lyr)\\    
HD 35601 & P                    & HD 181615  & EmL (Ca)    \\     
HD 42475 & P                    & HD 217476  & Var         \\     
HD 46687 & C                    & HD 222107  & Var (RS CVn)\\    
HD 54300 & C                    & BD+ 61 154 & EmL (Ca,H)  \\
HD 58972 & SB                   & M5  IV-87  & *           \\ 
HD 60522 & Var                  & M92 I-13   & *           \\
HD 74000 & *                    & M92 II-23  & *           \\    
HD 112014& SB                   & M92 VI-74  & *           \\    
HD 115604& Var                  &            &             \\    
\hline
\end{tabular}            
}                                                                          
\end{table}

It should be noted that, after computing any local fit in the above
procedure, we made sure that the residuals did not present any
systematic deviation, especially for stars of a given cluster or
metallicity range. In some cases we found single stars with small
index errors deviating excessively from the local fit, raising the
residual variance of the final fit. When this occurred, the stars were
analyzed in detail (looking for uncertainties in the atmospheric
parameters, variability, chromospheric activity, spectroscopic
binarity, etc) and, when necessary, they were rejected from the
fit. These stars are listed in Table~\ref{rejected}. Furthermore, all
the previous steps in the fitting procedure (e.g. boxes design,
overlapping zones definition, etc.) were also optimized to enlarge the
quality and accuracy of the final fitting functions.

\subsection{Fitting functions for the indices CaT$^*$ and PaT}

\begin{figure}
\centerline{\hbox{
\psfig{figure=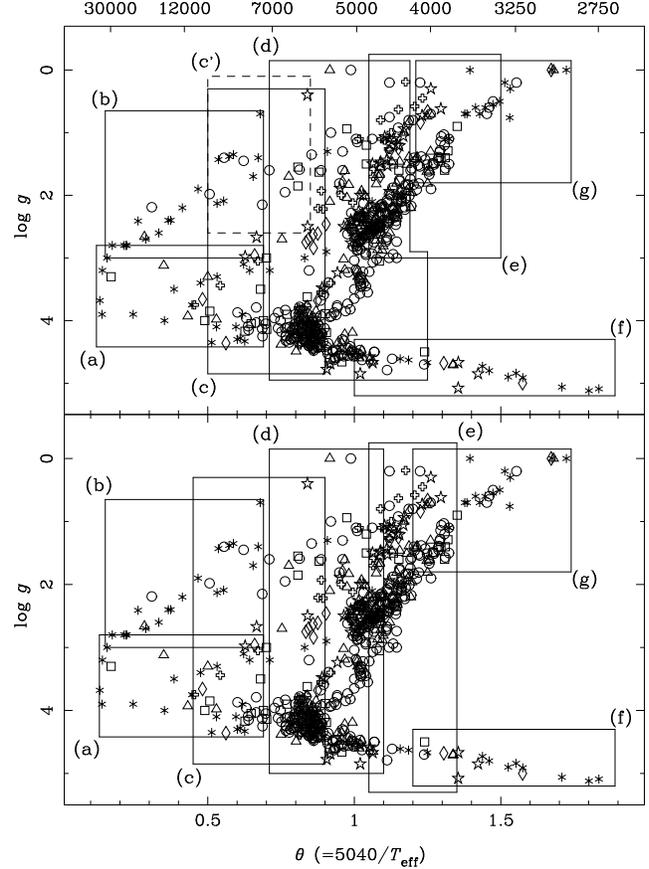}
}}  

\caption{$\log g$ -- $\theta$ diagrams for the whole stellar
library. The upper scale gives effective temperature in K.  Different
symbols are used to indicate stars of different metallicities, as
coded in Fig.~\ref{triplot}a. The boxes (labelled as in
Tables~\ref{coeffsCaT} and \ref{coeffsPaT}) display the regions of the
corresponding local fitting functions for the CaT$^{*}$ (upper) and
PaT (lower) indices.  In the overlapping zones, the final fitting
functions were derived by interpolating between boxes, as described in
the text. Except for the box (c') in the upper diagram, specially
designed to reproduce the proper behaviour of warm giants with low
metallicities (dashed line), the rest of the regions are defined for
the whole range of metallicities.}
\label{cajas}
\end{figure}

The derived local fitting functions for the indices CaT$^*$ and PaT
are presented, respectively, in Tables~\ref{coeffsCaT} and
\ref{coeffsPaT}. The Tables are subdivided according to the
atmospheric parameters ranges of each fitting box and include: the
functional forms of the fits (polynomial or exponential), the
significant coefficients and their corresponding formal errors, the
typical index error for the $N$ stars employed in each interval
($\sigma^{2}_{\rm typ}=N/\sum_{i=1}^{N}\sigma_{i}^{-2}$), the unbiased
residual variance of the fit ($\sigma^{2}_{\rm std}$) and the
determination coefficient ($r^{2}$). Note that this coefficient
provides the fraction of the index variation in the sample which is
explained by the derived fitting functions. Figure~\ref{cajas}
displays the parameter regions (labelled according to the atmospheric
parameters regimes of Tables~\ref{coeffsCaT} and \ref{coeffsPaT}) in
which the final fitting function was calculated by interpolating
between the overlapping zones. Note that, in some cases, these boxes
are somewhat narrower that the fitting regions listed in
Tables~\ref{coeffsCaT} and \ref{coeffsPaT}. This guarantees smoother
transitions between different regimes.

Users interested in implementing these fitting functions into their
population synthesis codes can make use of the {\sc fortran}
subroutine included in the URL addresses given above. This program
performs the required interpolations to provide the CaT$^*$, PaT and
CaT (computed as CaT $= {\rm CaT}^* + 0.93{\rm PaT}$) indices as a
function of the three input atmospheric parameters. It also gives an
estimation of the errors in the predicted indices (see below).

\begin{scriptsize}
\begin{table}                         
\caption{Coefficients and statistical data of the local fitting
functions for the index CaT$^*$ in each range of atmospheric
parameters. The term ``giant'' also includes supergiant stars.}
\begin{tabular}{l@{}lr@{}c@{}ll}          
\hline\hline    
(a) Hot dwarfs             & &   \multicolumn{3}{c}{0.13 $<$ $\theta$ $<$ 0.69}    & 2.75 $<$ $\log$ $g$ $<$ 4.37          \\
\hline
polynomial fit             & &         &     &                                     & $N$ = 44                              \\
$c_{0}$                    &:&    1.010&~$\pm$&~0.333                              & $\sigma_{{\rm typ}}$ = 0.193 \\
$\theta^{2}$               &:&  --40.80&~$\pm$&~5.24                               & $\sigma_{{\rm std}}$ = 0.641          \\
$\theta^{3}$               &:&    67.12&~$\pm$&~7.29                               & $r^{2}$ = 0.85                        \\
\hline\hline    
(b) Hot giants             & &   \multicolumn{3}{c}{0.14 $<$ $\theta$ $<$ 0.69}    & 0.69 $<$ $\log$ $g$ $<$ 3.01          \\
\hline
polynomial fit             & &         &     &                                     & $N$ = 26                              \\
$c_{0}$                    &:& --0.4492&~$\pm$&~0.2930                             & $\sigma_{{\rm typ}}$ = 0.148 \\
$\theta^{3}$               &:&    20.46&~$\pm$&~1.64                               & $\sigma_{{\rm std}}$ = 0.612          \\
                           & &         &     &                                     & $r^{2}$ = 0.93                        \\
\hline\hline    
(c) Warm stars             & &   \multicolumn{3}{c}{0.50 $<$ $\theta$ $<$ 0.90}    & 0.40 $<$ $\log$ $g$ $<$ 4.53          \\
\hline
polynomial fit             & &         &     &                                     & $N$ = 193                             \\
$c_{0}$                    &:& --27.12 &~$\pm$&~5.62                               & $\sigma_{{\rm typ}}$ = 0.192 \\
$\log$ $g$                 &:&   18.27 &~$\pm$&~5.69                               & $\sigma_{{\rm std}}$ = 0.649          \\
$\theta$~[Fe/H]            &:&   36.08 &~$\pm$&~6.16                               & $r^{2}$ = 0.94                        \\
$\theta$~$\log$ $g$        &:& --29.44 &~$\pm$&~7.25                               &                                       \\
$\theta^{2}$               &:&   127.8 &~$\pm$&~17.2                               &                                       \\
$\log^{2}$$g$              &:& --2.932 &~$\pm$&~0.983                              &                                       \\
${\rm [Fe/H]}^{2}$         &:&   3.611 &~$\pm$&~1.256                              &                                       \\
$\log$ $g$~[Fe/H]          &:& --5.038 &~$\pm$&~1.051                              &                                       \\ 
$\theta^{3}$               &:& --73.18 &~$\pm$&~10.86                              &                                       \\
$\log^{2}$$g$~[Fe/H]       &:&  0.4446 &~$\pm$&~0.1701                             &                                       \\
$\theta^{2}$~[Fe/H]        &:& --20.51 &~$\pm$&~4.69                               &                                       \\
$\theta$~$\log^{2}$$g$     &:&   4.406 &~$\pm$&~1.239                              &                                       \\
$\log$ $g$~[Fe/H]$^{2}$    &:&--0.8168 &~$\pm$&~0.3179                             &                                       \\
\hline\hline    
(c') Warm giants           & &   \multicolumn{3}{c}{0.13 $<$ $\theta$ $<$ 1.10}    & 0.10 $<$ $\log$ $g$ $<$ 3.10          \\
\ \ \ \ \ \ metal--poor    & &         &     &                                     & [Fe/H] $<$--0.25                      \\
\hline
polynomial fit             & &         &     &                                     & $N$ = 85                              \\
$c_{0}$                    &:&--0.1213 &~$\pm$&~2.277                              & $\sigma_{{\rm typ}}$ = 0.230 \\
$\theta$                   &:&   8.780 &~$\pm$&~2.056                              & $\sigma_{{\rm std}}$ = 0.637          \\
$\log$ $g$                 &:&--0.3118 &~$\pm$&~0.4345                             & $r^{2}$ = 0.89                        \\
$\theta$~[Fe/H]            &:&   2.396 &~$\pm$&~0.337                              &                                       \\
\hline\hline    
(d) Cool stars             & &   \multicolumn{3}{c}{0.70 $<$ $\theta$ $<$ 1.30}    & 0.00 $<$ $\log$ $g$ $<$ 4.85          \\
\hline
polynomial fit             & &         &     &                                     & $N$ = 551                             \\
$c_{0}$                    &:& --70.87 &~$\pm$&~14.56                              & $\sigma_{{\rm typ}}$ = 0.191 \\   
$\theta$                   &:&   302.2 &~$\pm$&~44.0                               & $\sigma_{{\rm std}}$ = 0.540          \\
$\log$ $g$                 &:& --20.58 &~$\pm$&~2.11                               & $r^{2}$ = 0.95                        \\
${\rm [Fe/H]}$             &:&   57.60 &~$\pm$&~11.20                              &                                       \\
$\theta$~[Fe/H]            &:& --80.81 &~$\pm$&~20.67                              &                                       \\
$\theta$~$\log$ $g$        &:&   12.77 &~$\pm$&~1.88                               &                                       \\
$\theta^{2}$               &:& --312.5 &~$\pm$&~44.2                               &                                       \\
$\log^{2}$$g$              &:&   3.514 &~$\pm$&~0.453                              &                               	   \\
${\rm [Fe/H]}^{2}$         &:&   2.314 &~$\pm$&~0.888                              &                               	   \\
$\log$ $g$~[Fe/H]          &:& --5.789 &~$\pm$&~0.971                              &                               	   \\
$\theta^{3}$               &:&   99.75 &~$\pm$&~14.51                              &                               	   \\
$\log^{3}$$g$              &:&--0.1520 &~$\pm$&~0.0323                             &                               	   \\
$\log^{2}$$g$~[Fe/H]       &:&  0.1315 &~$\pm$&~0.0738                             &                               	   \\
$\theta^{2}$~[Fe/H]        &:&   29.58 &~$\pm$&~9.29                               &                               	   \\
$\theta$~${\rm [Fe/H]}^{2}$&:& --2.103 &~$\pm$&~0.916                              &                               	   \\
$\theta$~$\log^{2}$$g$     &:& --1.674 &~$\pm$&~0.329                              &                               	   \\
$\theta$~$\log$ $g$~[Fe/H] &:&   4.071 &~$\pm$&~0.886                              &                               	   \\
\hline
\end{tabular}                                                                   
\label{coeffsCaT}
\end{table}   

\begin{table}                         
\contcaption{}
\begin{tabular}{l@{}lr@{}c@{}ll}          
\hline\hline    
(e) Cool giants            & &   \multicolumn{3}{c}{1.00 $<$ $\theta$ $<$ 1.40}    & 0.00 $<$ $\log$ $g$ $<$ 3.50          \\
\hline
polynomial fit             & &         &     &                                     & $N$ = 287                             \\
$c_{0}$                    &:&   368.9 &~$\pm$&~146.4                              & $\sigma_{{\rm typ}}$ = 0.182 \\   
$\theta$                   &:& --884.9 &~$\pm$&~382.9                              & $\sigma_{{\rm std}}$ = 0.529          \\
$\log$ $g$                 &:& --7.260 &~$\pm$&~2.101                              & $r^{2}$ = 0.91                        \\
${\rm [Fe/H]}$             &:&   10.40 &~$\pm$&~2.96                               &                                       \\
$\theta$~[Fe/H]            &:& --5.575 &~$\pm$&~2.477                              &                                       \\
$\theta^{2}$               &:&   742.7 &~$\pm$&~333.4                              &                                       \\
$\log^{2}$$g$              &:&  0.6285 &~$\pm$&~0.1689                             &                               	   \\
$\log$ $g$~[Fe/H]          &:&--0.8552 &~$\pm$&~0.2305                             &                               	   \\
$\theta^{2}$~$\log$ $g$    &:&   2.389 &~$\pm$&~1.312                              &                               	   \\
$\theta^{3}$               &:& --209.8 &~$\pm$&~96.4                               &                               	   \\
\hline\hline    
(f) Cold dwarfs            & &   \multicolumn{3}{c}{1.07 $<$ $\theta$ $<$ 1.84}    & 4.45 $<$ $\log$ $g$ $<$ 5.13          \\
\hline
polynomial fit             & &         &     &                                     & $N$ = 23                              \\
$c_{0}$                    &:& --104.7 &~$\pm$&~23.8                               & $\sigma_{{\rm typ}}$ = 0.200 \\
$\theta$                   &:&   245.1 &~$\pm$&~51.1                               & $\sigma_{{\rm std}}$ = 0.290          \\
$\theta^{2}$               &:& --173.2 &~$\pm$&~36.0                               & $r^{2}$ = 0.99                        \\
$\theta^{3}$               &:&   38.71 &~$\pm$&~8.35                               &                                       \\
\hline\hline    
(g) Cold giants            & &   \multicolumn{3}{c}{1.30 $<$ $\theta$ $<$ 1.73}    & 0.00 $<$ $\log$ $g$ $<$ 1.60          \\
\hline
exponential fit            & &    cte. & ~ =  &~2.0                                & $N$ = 27                              \\
$c_{0}$                    &:& --29.66 &~$\pm$&~14.28                              & $\sigma_{{\rm typ}}$ = 0.127 \\
$\theta$                   &:&   48.10 &~$\pm$&~19.76                              & $\sigma_{{\rm std}}$ = 0.888          \\
$\theta^{2}$               &:& --18.21 &~$\pm$&~6.80                               & $r^{2}$ = 0.83                        \\
\hline
\end{tabular}                                                                   
\end{table}   
\end{scriptsize}

\begin{scriptsize}
\begin{table}                         
\caption{Coefficients and statistical data of the local fitting
functions for the index PaT in each range of atmospheric
parameters. The term ``giant'' also includes supergiant stars.}
\begin{tabular}{l@{}lr@{}c@{}ll}          
\hline\hline    
(a) Hot dwarfs             & &   \multicolumn{3}{c}{0.13 $<$ $\theta$ $<$ 0.69}    & 2.75 $<$ $\log$ $g$ $<$ 4.37          \\
\hline
polynomial fit             & &         &     &                                     & $N$ = 46                              \\
$c_{0}$                    &:&  --6.415&~$\pm$&~0.792                              & $\sigma_{{\rm typ}}$ = 0.165 \\
$\theta$                   &:&    54.71&~$\pm$&~3.39                               & $\sigma_{{\rm std}}$ = 0.911          \\
$\theta^{2}$~$\log$ $g$    &:&  --4.283&~$\pm$&~1.091                              & $r^{2}$ = 0.91                        \\
$\theta^{3}$               &:&  --55.44&~$\pm$&~5.66                               &                                       \\
\hline\hline    
(b) Hot giants             & &   \multicolumn{3}{c}{0.14 $<$ $\theta$ $<$ 0.69}    & 0.69 $<$ $\log$ $g$ $<$ 3.01          \\
\hline
polynomial fit             & &         &     &                                     & $N$ = 29                              \\
$c_{0}$                    &:&    1.601&~$\pm$&~1.512                              & $\sigma_{{\rm typ}}$ = 0.123 \\
$\theta^{2}$               &:&    44.81&~$\pm$&~21.41                              & $\sigma_{{\rm std}}$ = 1.310          \\
$\theta^{3}$               & &  --46.88&~$\pm$&~29.20                              & $r^{2}$ = 0.64                        \\
\hline\hline    
(c) Warm stars             & &   \multicolumn{3}{c}{0.50 $<$ $\theta$ $<$ 0.90}    & 0.40 $<$ $\log$ $g$ $<$ 4.53          \\
\hline
polynomial fit             & &         &     &                                     & $N$ = 193                             \\
$c_{0}$                    &:& --149.0 &~$\pm$&~21.7                               & $\sigma_{{\rm typ}}$ = 0.170 \\
$\theta$                   &:&   651.2 &~$\pm$&~88.4                               & $\sigma_{{\rm std}}$ = 0.553          \\
$\log$ $g$                 &:&   18.74 &~$\pm$&~3.83                               & $r^{2}$ = 0.96                        \\
$\theta$~[Fe/H]            &:& --4.651 &~$\pm$&~1.766                              &                                       \\
$\theta$~$\log$ $g$        &:& --56.40 &~$\pm$&~10.92                              &                                       \\
$\theta^{2}$               &:& --840.9 &~$\pm$&~124.0                              &                                       \\
$\theta^{2}$~$\log$ $g$    &:&   39.26 &~$\pm$&~7.62                               &                                       \\
$\theta^{3}$               &:&   338.0 &~$\pm$&~60.0                               &                                       \\
$\theta^{2}$~[Fe/H]        &:&   5.880 &~$\pm$&~2.139                              &                                       \\
\hline\hline    
(d) Cool stars             & &   \multicolumn{3}{c}{0.70 $<$ $\theta$ $<$ 1.30}    & 0.00 $<$ $\log$ $g$ $<$ 4.85          \\
\hline
polynomial fit             & &         &     &                                     & $N$ = 551                             \\
$c_{0}$                    &:&   177.1 &~$\pm$&~12.0                               & $\sigma_{{\rm typ}}$ = 0.171 \\   
$\theta$                   &:& --445.5 &~$\pm$&~33.8                               & $\sigma_{{\rm std}}$ = 0.296          \\
$\log$ $g$                 &:& --16.00 &~$\pm$&~1.38                               & $r^{2}$ = 0.89                        \\
$\theta$~$\log$ $g$        &:&   29.20 &~$\pm$&~2.82                               &                                       \\
$\theta^{2}$               &:&   372.9 &~$\pm$&~31.2                               &                                       \\
${\rm [Fe/H]}^{2}$         &:&--0.3610 &~$\pm$&~0.1033                             &                               	   \\
$\theta^{2}$~$\log$ $g$    &:& --13.33 &~$\pm$&~1.43                               &                                       \\
$\theta^{3}$               &:& --103.3 &~$\pm$&~9.50                               &                               	   \\
${\rm [Fe/H]}^{3}$         &:&--0.1080 &~$\pm$&~0.0453                             &                               	   \\
\hline\hline    
(e) Very cool stars        & &   \multicolumn{3}{c}{1.00 $<$ $\theta$ $<$ 1.30}    & 0.00 $<$ $\log$ $g$ $<$ 4.85          \\
\hline
polynomial fit             & &         &     &                                     & $N$ = 300                             \\
$c_{0}$                    &:&  0.7600 &~$\pm$&~0.0343                             & $\sigma_{{\rm typ}}$ = 0.166 \\   
${\rm [Fe/H]}$             &:&  0.1743 &~$\pm$&~0.0550                             & $\sigma_{{\rm std}}$ = 0.251          \\
                           &:&         &     &                                     & $r^{2}$ = 0.13                        \\
\hline\hline    
(f) Cold dwarfs            & &   \multicolumn{3}{c}{1.07 $<$ $\theta$ $<$ 1.84}    & 4.45 $<$ $\log$ $g$ $<$ 5.13          \\
\hline
polynomial fit             & &         &     &                                     & $N$ = 23                              \\
$c_{0}$                    &:&   1.200 &~$\pm$&~0.208                              & $\sigma_{{\rm typ}}$ = 0.177 \\
$\theta^{3}$               &:&--0.3094 &~$\pm$&~0.0679                             & $\sigma_{{\rm std}}$ = 0.374          \\
                           &:&         &     &                                     & $r^{2}$ = 0.63                        \\
\hline\hline    
(g) Cold giants            & &   \multicolumn{3}{c}{1.25 $<$ $\theta$ $<$ 1.73}    & 0.00 $<$ $\log$ $g$ $<$ 1.65          \\
\hline
polynomial fit             & &         &     &                                     & $N$ = 44                              \\
$c_{0}$                    &:&   344.3 &~$\pm$&~183.8                              & $\sigma_{{\rm typ}}$ = 0.112 \\
$\theta$                   &:& --740.2 &~$\pm$&~374.5                              & $\sigma_{{\rm std}}$ = 0.351          \\
$\theta^{2}$               &:&   527.0 &~$\pm$&~253.0                              & $r^{2}$ = 0.79                        \\
$\theta^{3}$               &:& --123.8 &~$\pm$&~56.7                               &                                       \\
\hline
\end{tabular}                                                                   
\label{coeffsPaT}
\end{table}   
\end{scriptsize}

\subsubsection{CaT$^*$ fitting functions}

\begin{figure*}
\centerline{\hbox{
\psfig{figure=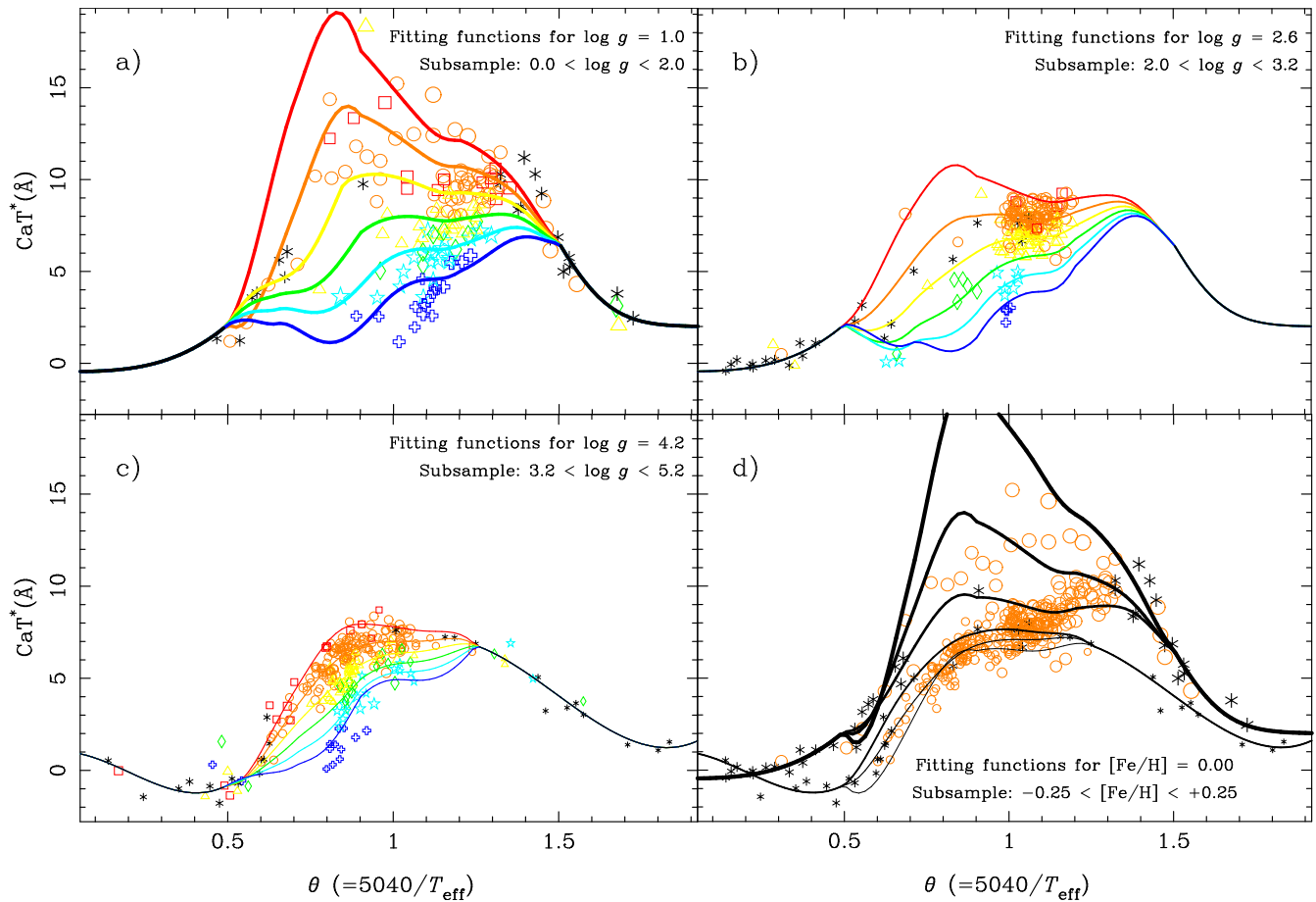}
}}                                                                                              
\caption{CaT$^*$ values and general fitting functions for different
atmospheric parameter regimes. Panels $a$, $b$ and $c$ display,
respectively, all the stars with gravities in the ranges $0.0\leq\log
g<2.0$, $2.0\leq\log g<3.2$ and $3.2\leq\log g<5.2$, together with the
derived fitting functions for the mean gravity in each range, i.e.,
$\log g =$ 1.0 (a), 2.6 (b) and 4.2 (c). In the mid-temperature range,
the different lines represent, from top to bottom, the functions for
metallicities [Fe/H] = +0.5, 0.0, --0.5, --1.0, --1.5 and --2.0,
whilst, for high and low temperatures, the fitting functions do not
depend on metallicity. Panel $d$ shows all the stars around solar
metallicity ($-0.25<{\rm [Fe/H]}\leq+0.25$) and the corresponding
fitting functions computed for [Fe/H] = 0.0 and different values of
gravity ($\log g$ = 0.0, 1.0, 2.0, 3.0, 4.0 and 5.0, from the thickest
to the thinnest line). Codes and relative sizes of the star symbols
(indicating, respectively, metallicity and gravity ranges) are
explained in Fig.~\ref{triplot}a. Note that, while the lines displayed
here correspond to fitting functions at particular values of $\log g$
and [Fe/H], the plotted stars span a range of atmospheric parameters
around these central values.  This is the reason why the lines do not
exactly fit all the points in the plots. Also, note that these fitting
functions have not been derived by using only these plotted stars, but
the whole sample.}
\label{fitCaTPaT}
\end{figure*}

As it is shown in Fig.~\ref{triplot}c, the CaT$^*$ index in hot stars
($0.13 \la \theta \la 0.69$) shows a dichotomic trend for dwarfs on
one side, and giants and supergiants on the other one. As a result of
this, we designed boxes a and b in Fig.~\ref{cajas} (top) and derived
different local fitting functions for the two groups of stars. Gravity
effects are nearly negligible within each of the two luminosity
bins. Also, due to the expected independence on metallicity for high
temperature stars, only terms in $\theta$ were found to be
statistically significant to reproduce the increasing CaT$^*$ trend
with decreasing temperature (Tables~\ref{coeffsCaT}a and
\ref{coeffsCaT}b). In the low temperature regime, the behaviour of
cold dwarfs also differs from that of cold giants
(Fig.~\ref{triplot}c). Here, we followed the same strategy and derived
two different local fits in which, again, only terms in $\theta$ were
needed (Tables~\ref{coeffsCaT}f and \ref{coeffsCaT}g). As we mentioned
in Section~\ref{qualitative}, the decreasing CaT$^*$ trend with
increasing $\theta$ starts at lower temperatures for giants ($\theta
\simeq 1.30$) than for dwarfs ($\theta \simeq 1.07$), being also
steeper for the former group of stars. Finally, it is worth noting
that, although a considerable number of hot and cold stars have
unknown [Fe/H] determinations, the small scatter in the data shows
that metallicity is indeed not governing the observed CaT$^*$
behaviour at these temperature regimes.

The complex behaviour of warm and cool stars was parametrized in terms
of the three atmospheric parameters (see Tables~\ref{coeffsCaT}c,
\ref{coeffsCaT}d and \ref{coeffsCaT}e). Although an unique parameter
box enclosing the rest of stars was enough to reproduce the CaT$^*$
behaviour at intermediate temperatures (box d in upper panel of
Fig.~\ref{cajas}), we designed transition boxes (c and e) at both
sides to improve the quality of the interpolations with the local
functions derived for hot and cold giants stars. Note that cold dwarfs
follow a trend smoother than cold giants, and we did not need to
include them in box d. Figures~\ref{fitCaTPaT}a, \ref{fitCaTPaT}b and
\ref{fitCaTPaT}c illustrate the general fitting functions for three
gravity bins (representing, roughly, supergiants, giants and dwarfs),
showing an increasing dependence on metallicity as gravity decreases
(therefore, supergiants span the widest range in CaT$^*$
values). Figure~\ref{fitCaTPaT}d displays the derived fitting
functions for solar metallicity and different gravity values. It is
clear from this figure that gravity effects increase with decreasing
gravity.  Also, the higher the metallicity, the larger the gravity
dependence. These effects were already apparent in Figs.~\ref{fig_g}
and~\ref{fig_z}.

Due to the lack of warm metal-poor giants (there are only a few stars
in the sample), the local fit in Table~\ref{coeffsCaT}c' was specially
designed to avoid unreliable interpolations between hot and warm
giants in the low metallicity interval. Moreover, in order to ensure a
smooth interpolation, the limits of this box are slightly different
depending on the range of metallicity ($0.65 < \theta < 0.95$ for
${\rm [Fe/H]} \leq -1.25$, $0.65 < \theta < 0.90$ for 
$-1.25 < {\rm [Fe/H]} \leq -0.75$, and 
$0.65 < \theta < 0.85$ for $-0.75 < {\rm [Fe/H]} \leq -0.25$). 
Therefore, Fig.~\ref{cajas} only illustrates one particular case
(dashed line).

It is important to note that, while the lines in
Figure~\ref{fitCaTPaT} correspond to projections of the fitting
functions at particular values of $\log g$ and [Fe/H], the plotted
stars span a range of atmospheric parameters around these central
values. Therefore, it is not expected that the lines fit all the
points in the plots. However, there are still some curves which do not
appear to be well constrained by the observations in certain regions
of the parameter space. For instance, there are no stars for $\theta <
0.5$ in Figure~\ref{fitCaTPaT}a ($\log g = 1.0$) and for $\theta >
1.3$ in Figure~\ref{fitCaTPaT}b ($\log g = 2.6$). Stars with these
parameters do not exist (see Fig.~\ref{cajas}), and therefore they
will never be required by the stellar populations models. The curves
at [Fe/H]$ = + 0.5$ (Figure~\ref{fitCaTPaT}a) and at $\log g = 0.0$
(Figure~\ref{fitCaTPaT}d) for $0.50 < \theta < 1.0$ deserve a word of
caution. The problem here is that high abundance, relatively hot
supergiants are indeed rare and, therefore, uncertainties in the
CaT$^{*}$ predictions for this region may be present. Note that, since
the fitting functions have been derived using the whole sample of
stars within each parameter box, the CaT$^{*}$ predictions for high
metallicity, hot supergiants are partly driven by extrapolations of
the functional dependence determined from stars of lower abundances
and higher gravities. The contribution of this kind of stars (hot
supergiants with metallicities above solar) at the near-IR spectra of
old galaxies is not important, although it would not be the case for
very young stellar populations.

\subsubsection{PaT fitting functions}

\begin{figure}
\centerline{\hbox{
\psfig{figure=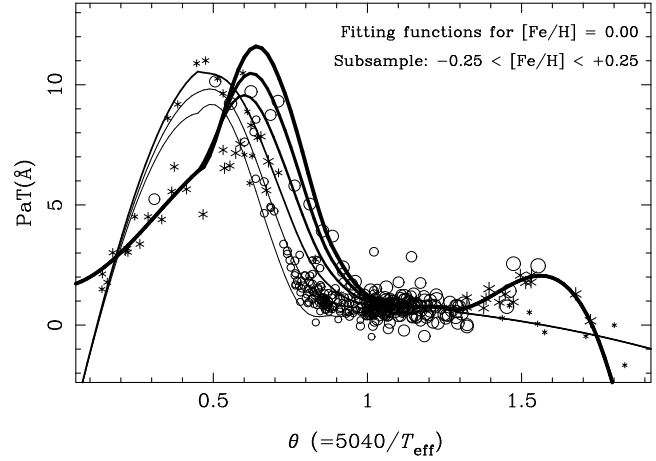}
}}                                                                                              
\caption{PaT values for all the stars around solar metallicity
($-0.25<{\rm [Fe/H]}\leq+0.25$). The curves correspond to general
fitting functions computed for [Fe/H] = 0.0 and different values of
gravity ($\log g$ = 0.0, 1.0, 2.0, 3.0, 4.0 and 5.0, from the thickest
to the thinnest line). Codes and relative sizes of the star symbols
(indicating, respectively, metallicity and gravity ranges) are
explained in Fig.~\ref{triplot}a.}

\label{fitPaT}
\end{figure}

As in the case of the CaT$^*$, the PaT index in hot and cold stars
shows a different trend for dwarfs and giants. Once again, we designed
proper boxes for each group (a, b, f and g in the lower panel of
Fig.~\ref{cajas}) and derived the local fitting functions given in
Tables~\ref{coeffsPaT}a, \ref{coeffsPaT}b, \ref{coeffsPaT}f and
\ref{coeffsPaT}g. Figure~\ref{fitPaT} shows the general fitting
functions for solar metallicity and different gravities. For hot
stars, PaT increases as temperature decreases, reaching a maximum at a
temperature that depends on the luminosity class. To reproduce the
dependence on gravity of the hot dwarfs, a term in log g was deamed
necessary.  At the cold temperature end, the PaT in giant stars
exhibits a bump due to the presence of strong TiO absorptions falling
into the index bandpasses.

Warm stars follow a decreasing trend with increasing $\theta$ up to
the cool stars which, as expected, attain values around zero. Boxes c
and d were mainly designed to reproduce the maximum and the above
trends. For cool stars, we have found a mild metallicity dependence in
the sense that the larger the metallicity the stronger the
PaT. Finally, box e allows smooth interpolations with the coldest
stars.

\subsection{Residuals and error analysis}

\begin{figure}
\centerline{\hbox{
\psfig{figure=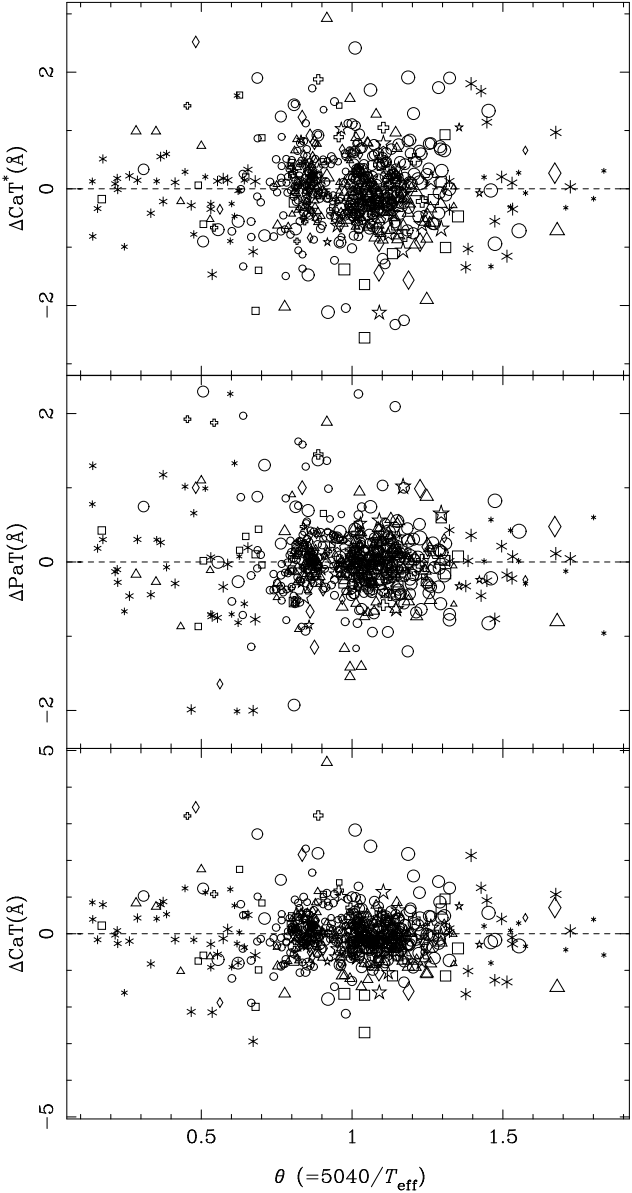}
}}
\caption{Residuals of the derived fitting functions ($\Delta {\cal
I} = {\cal I}_{\rm obs} - {\cal I}_{\rm pred}$) against $\theta$. See
Fig.~\ref{triplot}a for symbol codes.}
\label{residuals}
\end{figure}

We define the residual of any star as the difference between the
observed index and the one predicted by the fitting functions ($\Delta
{\cal I} = {\cal I}_{\rm obs} - {\cal I}_{\rm
pred}$). Figure~\ref{residuals} shows the residuals of the indices
CaT$^*$, PaT and CaT for the whole stellar library as a function of
$\theta$. The residuals do not exhibit systematic trends with any of
the three atmospheric parameters. Star clusters have also been
analyzed separately and, except for the globular cluster M71, no
systematic effects have been found. Compared to the predictions of the
fitting functions, this cluster shows a significant mean offset of
$\Delta {\rm CaT}^* = -0.35$ and $\Delta {\rm CaT} = -0.51$. The most
plausible source of these systematic offsets is an error in the
atmospheric parameters. A systematic error in the controversial
reddening of this cluster would affect the adopted values. However,
given the Ca\,{\sc ii} triplet dependence on atmospheric parameters in
this region of the parameters space, only a change in the adopted
metallicity could account for the derived offsets. The controversy
about the metallicity of M71 has already been discussed in Section~6.1
of Paper~II. In fact, these residuals were used as one of the
arguments in favour of the Carretta \& Gratton (1997) metallicity
scale (note that the value of ${\rm [Fe/H]}=-0.56$ established by the
Zinn \& West (1984) scale leads to even larger residuals). According
to the predictions of the new fitting functions, M71 should have a
metallicity of [Fe/H]$=-0.84\pm0.06$, instead of the adopted value of
${\rm [Fe/H]}=-0.70$.  Nevertheless, we have decided not to change the
input metallicity of this cluster, keeping the Carreta \& Gratton
scale for all the globular clusters. Note that the CaT metallicity
dependence is constrained by using more than 500 field and cluster
stars, and a change in the adopted metallicity of M71 would not
significantly alter the results. An anomalous [Ca/Fe] abundance ratio
for M71 could also explain the derived offsets. However, using the
tabulated data from Th\'evenin (1998), we derive a mean [Ca/Fe] ratio
for this cluster of $0.28 \pm 0.07$, which is significantly above the
expected mean [Ca/Fe] value (0.13) for the cluster metallicity (see
Figure~\ref{CaFe}) and, therefore, can not explain negative residuals
in the indices.

\begin{table}                         
\centering{                                                                
\caption{Statistical data for the general fitting functions of the
indices CaT$^*$, PaT and CaT. $\sigma_{\rm std}$: unbiased residual
standard deviation; $\sigma_{\rm typ}$: typical index error for the
stars used in the fits; $r^{2}$: determination coefficient.}
\begin{tabular}{lccc}
\hline
        & $\sigma_{\rm std}$ & $\sigma_{\rm typ}$ & $r^{2}$ \\
\hline
CaT$^*$ & 0.534              & 0.184              & 0.965   \\
PaT     & 0.427              & 0.163              & 0.965   \\ 
CaT     & 0.629              & 0.197              & 0.939   \\
\hline
\end{tabular}
\label{sigmas}
}
\end{table}

\begin{table*}                        
\centering{                                                                
\caption{Uncertainties of the CaT$^{*}$ fitting functions for different
groups of stars, and mean CaT$^{*}$ errors due to uncertainties in
the input atmospheric parameters. $N$: number of stars. $\sigma_{\rm
std}$: unbiased residual standard deviation of the fit. $\sigma_{\rm
typ}$: typical observational CaT$^{*}$ error for the subsample of
stars. $\sigma_{\rm res}$: residual error (derived as
$\sqrt{\sigma_{\rm std}^2-\sigma_{\rm typ}^2}$).  $\sigma_{T_{\rm
eff}}$, $\sigma_{\log g}$ and $\sigma_{\rm [Fe/H]}$: mean CaT$^{*}$
errors due to uncertainties in the input $T_{\rm eff}$, $\log g$ and
[Fe/H]. $\sigma_{{\rm par}}$: total error due to atmospheric
parameters (quadratic sum of the three previous errors).}
\begin{tabular}{lrccccccccc}
\hline
       & $N$ & & $\sigma_{\rm std}$ & $\sigma_{\rm typ}$ & $\sigma_{\rm res}$ & & $\sigma_{T_{\rm eff}}$ & $\sigma_{\log g}$ & $\sigma_{\rm [Fe/H]}$ & $\sigma_{{\rm par}}$ \\
\hline
Open clusters (Coma, Hyades, M67, NGC188, NGC7789)      &  93 & & 0.54 & 0.28 & 0.46 & & 0.24 & 0.07 & 0.42 & 0.49\\
Globular clusters (M3, M5, M10, M13, M71, M92, NGC6171) &  52 & & 0.84 & 0.76 & 0.36 & & 0.08 & 0.11 & 0.54 & 0.56\\
Field dwarfs                                            & 243 & & 0.45 & 0.18 & 0.41 & & 0.23 & 0.15 & 0.23 & 0.36\\
Field giants                                            & 201 & & 0.46 & 0.16 & 0.43 & & 0.15 & 0.26 & 0.23 & 0.38\\
Field supergiants                                       &  75 & & 0.84 & 0.17 & 0.82 & & 0.47 & 0.46 & 0.52 & 0.84\\
\hline		     					         		     
Hot stars ($0.13 < \theta < 0.69$)                      &  68 & & 0.66 & 0.17 & 0.64 & & 0.81 & 0.23 & 0.42 & 0.94\\
Intermediate stars ($0.69 < \theta < 1.30$)             & 555 & & 0.48 & 0.19 & 0.45 & & 0.12 & 0.19 & 0.31 & 0.38\\
Cold stars ($1.30 < \theta < 1.84$)                     &  41 & & 0.71 & 0.14 & 0.70 & & 0.69 & 0.20 & 0.21 & 0.75\\
\hline		     
All                                                     & 664 & & 0.53 & 0.18 & 0.50 & & 0.22 & 0.20 & 0.31 & 0.43\\
\hline
\end{tabular}
\label{errpredatm}
}
\end{table*}

In order to explore in more detail the reliability of the present
fitting functions, in Table~\ref{sigmas} we list the unbiased residual
standard deviation from the fits, $\sigma_{\rm std}$, the typical
error in the measured indices, $\sigma_{\rm typ}$, and the
determination coefficient, $r^{2}$, for the 664 stars employed in the
computation of the general fitting functions (note that, apart from
the 23 stars listed in Table~\ref{rejected}, 19 stars with unknown
[Fe/H] could not be included in the metallicity-dependent fits). It is
clear that, for the three indices, the computed $\sigma_{\rm std}$ is
larger than what it should be expected uniquely from index
uncertainties (see also the partial values of $\sigma_{\rm std}$ and
$\sigma_{\rm typ}$ in Tables~\ref{coeffsCaT}
and~\ref{coeffsPaT}). Since we are confident that the latter errors
are reliable (see full details on index errors computation in
Section~5 of Paper~I), the residuals must be dominated by a different
error source.  In fact, such an additional scatter could be due to
uncertainties in the input atmospheric parameters of the library
stars.  In order to check this, we have computed how the parameters
errors translate into uncertainties in the predicted CaT$^{*}$.  This
depends both on the local functional form of the fitting function
(e.g., a weak dependence on temperature leads to small index errors
due to $T_{\rm eff}$ uncertainties) and on the atmospheric parameters
range (e.g., hot stars have $T_{\rm eff}$ uncertainties larger than
cooler stars). Thus, we have carried out the following analysis for
different groups of stars (listed in Table~\ref{errpredatm}).

For each star of the sample we have derived three CaT$^{*}$ errors,
arising from the corresponding uncertainties in $T_{\rm eff}$, $\log
g$ and [Fe/H].  As input atmospheric parameters uncertainties we have
made use of the values presented in Table~7 of Paper~II. Apart from
those, we have used errors of 75~K, 0.40~dex and 0.15~dex for the
effective temperatures, gravities and metallicities taken from
Soubiran, Katz \& Cayrel (1998) (stars coded {\sc skc} in Table~6 of
Paper~II), and 75~K, 0.05~dex and 0.20~dex for the cluster stars.  For
each star group we have computed a mean CaT$^{*}$ error as a result of
the uncertainty of each parameter ($\sigma_{T_{\rm eff}}$,
$\sigma_{\log g}$ and $\sigma_{\rm [Fe/H]}$) by using the input
parameters errors for all the individual stars.  Finally, an estimate
of the total expected error due to atmospheric parameters
($\sigma_{{\rm par}}$) is computed as the quadratic sum of the three
previous errors.

Table~\ref{errpredatm} presents a comparison of these expected errors
with the unexplained residual errors of the fits ($\sigma_{\rm res}\equiv 
\sqrt{\sigma_{\rm std}^2-\sigma_{\rm typ}^2}$).
It is clear from this table that, in most cases, the additional
scatter ($\sigma_{\rm res}$) is comparable to $\sigma_{{\rm par}}$,
that is, it can be explained from uncertainties in the input
atmospheric parameters. Using an $F$ test of comparison of variances
we have checked that $\sigma_{\rm res}$ is not significantly larger
than $\sigma_{\rm par}$, except for the intermediate stars. This means
that, a minor, additional error source may still be needed to account
for the observed residuals of field dwarfs and giants with
intermediate temperatures.

\begin{table*}                         
\centering{                                                                
\caption{Absolute errors in the fitting functions
predictions for different values of the atmospheric parameters. Input
$\log g$ values varying with effective temperature for dwarfs, giants
and supergiants have been taken from Lang (1991). Since, for extreme
temperatures, the fitting functions do not depend on metallicity, no
[Fe/H] value has been adopted for 15000~K and 3200~K. This is also the
case for some values at 3500~K.}
\begin{tabular}{rrc@{}c@{}cc@{}c@{}cc@{}c@{}c}
\hline
    &
    &\multicolumn{3}{c}{dwarfs}&\multicolumn{3}{c}{giants}&\multicolumn{3}{c}{supergiants}\\
\hline
$T_{\rm eff}$\ \ &[Fe/H]& $\Delta$CaT*\ \ &$\Delta$PaT\ \ &$\Delta$CaT&$\Delta$CaT*\ \ &$\Delta$PaT\ \ &$\Delta$CaT&$\Delta$CaT*\ \ &$\Delta$PaT\ \ &$\Delta$CaT\\
\hline \medskip
15000&    &  0.17&  0.25& 0.29&  0.17&  0.25&  0.29&  0.24&  0.60&  0.61\\
8000& +0.5&  0.39&  0.29& 0.47&  0.45&  0.25&  0.51&  0.60&  0.32&  0.67\\
8000&  0.0&  0.28&  0.24& 0.36&  0.30&  0.20&  0.36&  0.29&  0.28&  0.39\\
8000&--1.0&  0.40&  0.33& 0.50&  0.37&  0.32&  0.47&  0.80&  0.39&  0.87\\ \medskip
8000&--2.0&  0.76&  0.54& 0.91&  0.77&  0.54&  0.92&  0.81&  0.59&  0.98\\
6000& +0.5&  0.25&  0.09& 0.26&  0.38&  0.11&  0.39&  0.60&  0.20&  0.63\\
6000&  0.0&  0.10&  0.06& 0.11&  0.22&  0.09&  0.23&  0.37&  0.20&  0.41\\
6000&--1.0&  0.17&  0.09& 0.19&  0.34&  0.11&  0.35&  0.50&  0.21&  0.54\\ \medskip
6000&--2.0&  0.37&  0.16& 0.40&  0.58&  0.18&  0.60&  0.68&  0.25&  0.72\\
5000& +0.5&  0.28&  0.07& 0.29&  0.19&  0.04&  0.20&  0.31&  0.08&  0.32\\
5000&  0.0&  0.17&  0.07& 0.18&  0.10&  0.04&  0.11&  0.18&  0.08&  0.20\\
5000&--1.0&  0.19&  0.08& 0.20&  0.14&  0.06&  0.15&  0.28&  0.09&  0.29\\ \medskip
5000&--2.0&  0.39&  0.12& 0.40&  0.29&  0.10&  0.31&  0.41&  0.11&  0.42\\
4000& +0.5&  0.11&  0.08& 0.13&  0.41&  0.13&  0.42&  0.66&  0.13&  0.68\\
4000&  0.0&  0.11&  0.07& 0.13&  0.26&  0.12&  0.28&  0.54&  0.12&  0.55\\
4000&--1.0&  0.11&  0.07& 0.13&  0.38&  0.13&  0.40&  0.40&  0.13&  0.42\\ \medskip
4000&--2.0&  0.11&  0.10& 0.15&  0.74&  0.16&  0.75&  0.54&  0.16&  0.56\\
3500& +0.5&  0.13&  0.10& 0.16&  0.97&  0.19&  0.98&  0.99&  0.19&  1.00\\
3500&  0.0&  0.13&  0.10& 0.16&  1.00&  0.19&  1.02&  1.00&  0.19&  1.01\\
3500&--1.0&  0.13&  0.10& 0.16&  1.13&  0.19&  1.14&  1.07&  0.19&  1.09\\ \medskip
3500&--2.0&  0.13&  0.10& 0.16&  1.30&  0.19&  1.31&  1.20&  0.19&  1.22\\
3200&     &  0.19&  0.13& 0.23&  0.37&  0.22&  0.42&  0.37&  0.22&  0.42\\
\hline
\end{tabular}
\label{errpred}
}
\end{table*}

Finally, since the aim of this paper is to predict reliable index
values for any combination of input atmospheric parameters, we have
also computed, making use of the covariance matrices of the fits, the
random errors in such predictions. These uncertainties are given in
Table~\ref{errpred} for some representative values of input
parameters. Note that, as is expected, the absolute errors are
larger for cold giants and supergiants and, in general, increase as
the metallicity departs from the solar value.

\subsection{Ca/Fe abundance ratios}
\label{CaFeratio}

\begin{figure}
\centerline{\hbox{
\psfig{figure=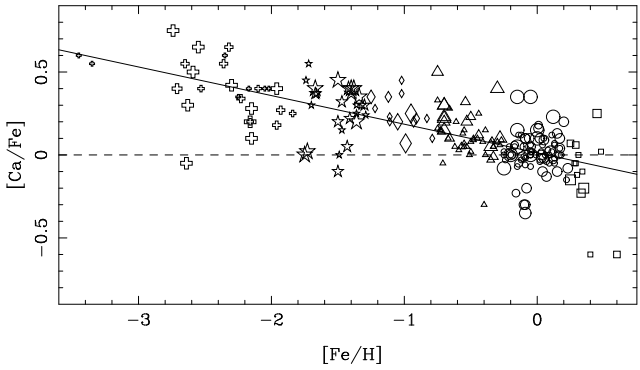}
}}
\caption{[Ca/Fe] ratios against [Fe/H] for 216 stars of the
library. Different symbols and sizes indicate metallicities and
gravities as in Fig.~\ref{triplot}a. The relationship ${\rm [Ca/Fe]}_{0} =
0.013(\pm0.012) - 0.173(\pm0.011)~{\rm [Fe/H]}$ (solid line) is a
least-squares fit to the
data and confirms an intrinsic anticorrelation between 
both abundance ratios.}
\label{CaFe}
\end{figure}

\begin{figure}
\centerline{\hbox{
\psfig{figure=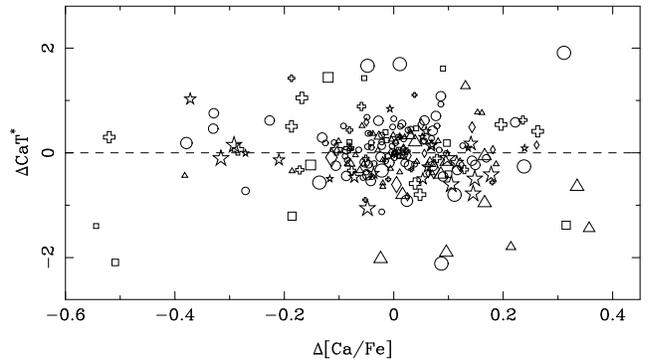}
}} 
\caption{Residuals of the CaT$^{*}$ fitting functions (observed --
predicted) versus residuals of the linear [Ca/Fe]
relationship. Symbols and sizes for the 216 stars are as in
Fig.~\ref{triplot}a.}
\label{resCaFe}
\end{figure}

Since CaT$^*$ is a pure Ca\,{\sc ii} index, the parameter governing
the metallicity dependence should be the actual [Ca/H] ratio, rather
than the classical [Fe/H]. Since, in spite of this, we have been using
[Fe/H] as our metallicity parameter, we have investigated whether the
residuals of the CaT$^*$ fitting functions are correlated with the
individual Ca/Fe abundance ratios. Note that, if that were the case,
this could account for part of the unexplained residuals reported in
the previous section.  Using a subsample of 216 library stars, for
which we have compiled [Ca/Fe] data from the literature (Gratton \&
Sneden 1987; Pilachowski, Sneden \& Kraft 1996; Nissen \& Schuster
1997; Th\'evenin 1998), we have not found any evidence for such an
expected correlation. This agrees with the results of ITD, which,
after including a [Ca/Fe] term in their fitting functions, concluded
that this relative abundance did not practically affect the value of
CaT. In fact, a proper error analysis demonstrates that their derived
[Ca/Fe] coefficient is not statistically different from zero.

To investigate this issue in more detail, in Figure~\ref{CaFe} we plot
the [Ca/Fe] relative abundances versus the assumed metallicities in
[Fe/H]. As it has been previously reported (e.g. Edvardsson et
al. 1993), there exists a clear anticorrelation between both abundance
ratios for Galactic stars. A least-squares linear fit to all the stars
in this plot gives ${\rm [Ca/Fe]}_{0} = 0.013(\pm0.012) -
0.173(\pm0.011)~{\rm [Fe/H]}$.  The existence of this relation
indicates that the actual calcium abundance has been implicitly taken
into account in the fitting functions through the adopted [Fe/H] and,
therefore, a systematic trend between the fitting functions residuals
and the Ca/Fe ratios should not be expected. However we could still
have a correlation between the residuals from the above relation
($\Delta {\rm [Ca/Fe]} = {\rm [Ca/Fe]} - {\rm [Ca/Fe]}_{0}$), and the
fitting functions residuals. Figure~\ref{resCaFe} shows that these
residuals are not correlated at all. Thus, we conclude that the new
fitting functions do not need any additional term to account for Ca
overabundances within the library stars. Note however, that the
fitting functions in their present form (i.e. using [Fe/H] as the
metallicity indicator) implicitly include the chemical enrichment of
the solar neighborhood, although, in principle, they could be
corrected, by introducing a new [Ca/Fe]--[Fe/H] relation, to give
predictions for other enrichment scenarios.

\section{Comparison with previous fitting functions}

In this section, we present a comparative analysis of the previous
fitting functions by DTT, ITD and JCJ (see Section~\ref{prevff} and
Table~\ref{prevcal}) and those presented in this paper. DTT and ITD
made use of empirical stellar libraries of 106 and 67 stars
repectively, and derived fitting functions for their calcium
indices. On the other hand, using NLTE models, JCJ derived fitting
functions for the theoretical equivalent widths of the Ca\,{\sc ii}
lines.

In order to avoid extrapolations of the previous fitting functions,
the range of effective temperatures in which comparisons have been
performed was constrained to the one approximately spanned by the
three previous works ($0.75 \leq \theta \leq 1.25$).  Also, since the
empirical papers by DTT and ITD do not correct for the Paschen
contamination, their predictions are compared with our CaT fitting
functions. For the comparison with JCJ we use the CaT$^{*}$ fitting
functions to compare with their theoretical Ca\,{\sc ii} predictions,
since they are based on model atmospheres which do not include the
contamination by other elements.

Prior to any comparison, the previous fitting functions have been
corrected from systematic offsets arising from differences in the
index definitions. For the predictions by DTT and ITD, we have applied
the calibrations given in Table~8 of Paper~I, which were specially
designed to convert the values of the literature indices to our
system. Although JCJ do not define line-strength indices, they provide
fitting functions for the total true equivalent width of the two
strongest Ca\,{\sc ii} lines. We have increased their predicted
equivalent widths by 21.1($\pm 0.3$) percent. Such a fraction, which
has been empirically derived using all the library stars within the
range of effective temperatures given above, quantifies the strength
of the weakest line of the triplet relative to the sum of the two
other ones. Hence, after these corrections, differences between the
predictions of the fitting functions will only arise from differences
in the input data quality (in the case of the empirical papers, this
will include the observational errors in the indices, the
uncertainties in the atmospheric parameters and the number of
calibrating stars), the range of atmospheric parameters spanned by the
library, and the mathematical procedure to compute the fits.

\begin{figure*}
\centerline{\hbox{
\psfig{figure=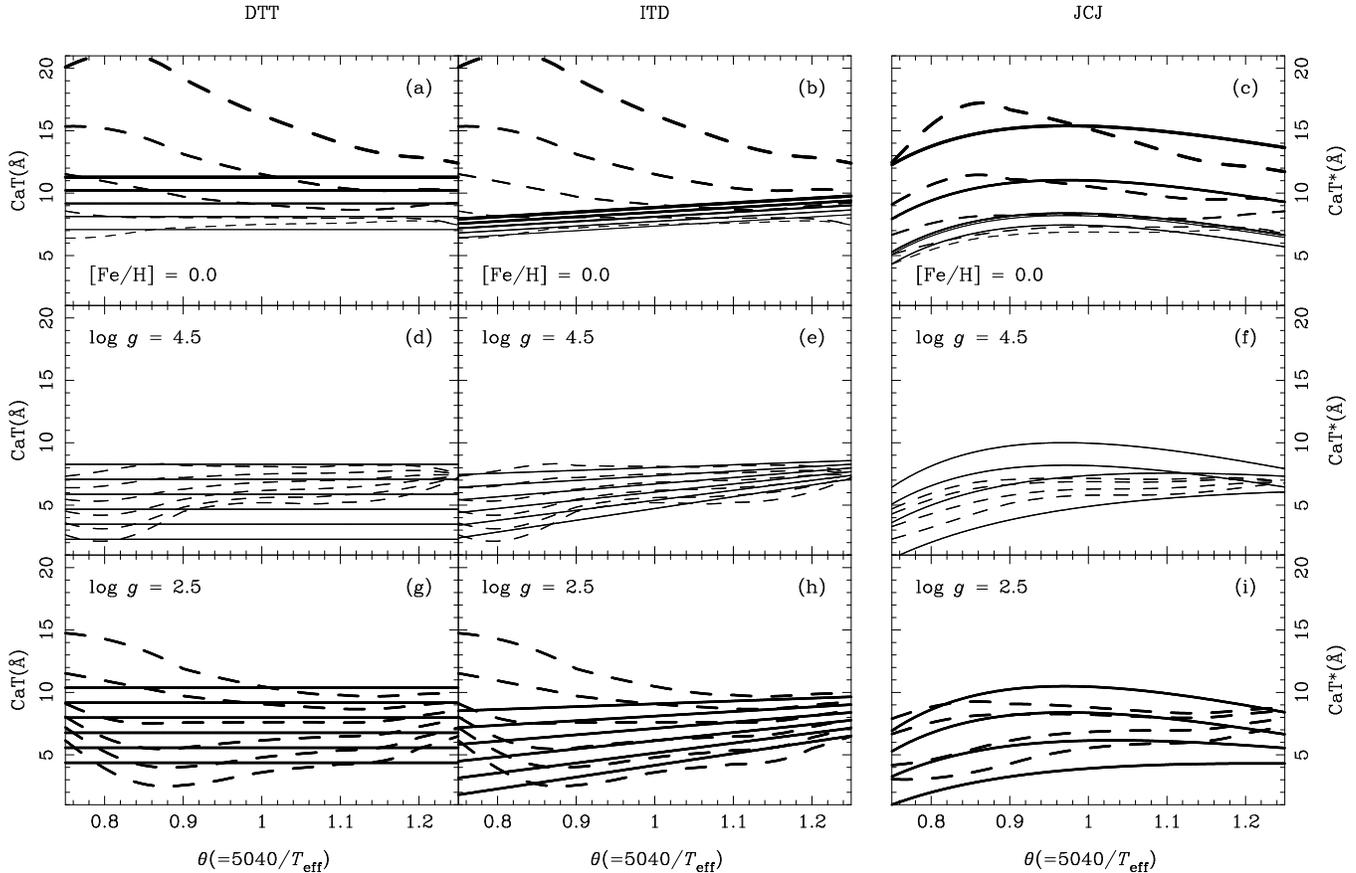}
}}  
\caption{Comparison with previous fitting functions. Panels a, b and c display,
respectively, the predictions by DTT, ITD and JCJ for solar
metallicity stars and different gravities ($\log g$ = 0.5, 1.5, 2.5,
3.5, and 4.5). The thicker the line, the lower the gravity. Solid
lines correspond to the fitting functions by the above authors, whilst
our predictions are given in dashed lines. Analogously, the predicted
dependence on metallicity is also compared for dwarfs ($\log g =
4.5$), in panels d, e and f, and giants ($\log g = 2.5$), in panels g,
h and i. The different lines represent, from top to bottom, the
metallicities [Fe/H] = +0.5, 0.0, --0.5, --1.0, --1.5, and --2.0, for
DTT and ITD, and [Fe/H] = +0.2, 0.0, --0.5, and --1.0 for JCJ. Note
that, in all cases, the predictions by DTT and ITD are compared with
the CaT fitting functions, whereas the theoretical work by JCJ is
compared with CaT$^*$.}
\label{compfit}
\end{figure*}

Figure~\ref{compfit} shows the present Ca\,{\sc ii} fitting
functions (dashed lines) together with those derived by DTT, ITD and
JCJ (solid lines). The predicted gravity effects are shown in the
upper panels (a, b, c), whereas mid (d, e, f) and lower panels (g, h,
i) show, respectively, the predicted dependences on metallicity
for dwarf and giants.

DTT found a biparametrical, linear Ca\,{\sc ii} behaviour with
metallicity and gravity, but no dependence on temperature (see also
the discussion in Section~\ref{prevff}). Concerning the gravity
effect, DTT predict a
milder dependence than what our data shows, 
as it is apparent from Fig.~\ref{compfit}a.
Also, there
are important differences at the other temperature ranges, although metallicity
effects agree with our predictions, especially for warm dwarfs
(Figs.~\ref{compfit}d) and cool giants (Figs.~\ref{compfit}g).  We must
note that most differences between DTT and our predictions arise from:
i) the lack of second order and $\log g$--[Fe/H] cross-terms in their
derived functions (a reanalysis of their data shows that these terms
were indeed statistically significant), and ii) the absence of an
effective temperature dependence. This last discrepancy is mainly due
the lack of cold ($\theta>1.05$) dwarfs in their sample (see the
effect in panel d) and to the contamination by Paschen lines (panel
g).

The fitting functions by ITD include terms in effective temperature
and several cross-terms in the parameters. They concluded that the
gravity effect is not as important as the dependence on
metallicity. Fig.~\ref{compfit}b shows that their linear, extremely
weak, gravity dependence, does not agree at all with our conclusions.
As it was discussed in Section~\ref{prevff}, this is mostly due to the
scarcity of supergiants in their sample and to the fact that most of
them have very low metallicities. On the other hand, the metallicity
dependence for dwarfs and cool giants roughly matches with our
predictions, although it is again very different for warm giants
(Figs.~\ref{compfit}e,h).

JCJ found a complex dependence on the three atmospheric
parameters. Unlike the two previous empirical works, there is a good
agreement between the qualitative behaviour of the parametrized
gravity dependence (Fig.~\ref{compfit}c), in the sense that the lower
the gravity, the larger the gravity effect. Also, they find a
metallicity dependence with a functional form similar to our
predictions, although their values are more sensitive to metallicity
than ours, especially for dwarfs (Figs.~\ref{compfit}f,i).

We conclude that, in the temperature range of comparisons, the gravity
dependence of the Ca\,{\sc ii} triplet predicted by JCJ, rather than
that of the empirical papers, roughly agrees with our results. On the
other hand, for certain temperature ranges, DTT and ITD better reproduce the
metallicity dependence found in this work.

\section{Conclusions}
\label{conclusions}

In this paper, we have analysed the behaviour of the Ca\,{\sc ii}
triplet strength in the spectra of stars of different spectral types
and luminosities, by means of a new near-IR stellar library (presented
in Paper~I) with a wide coverage of atmospheric parameters (Paper~II).
We have derived empirical fitting functions which can be easily
implemented into stellar populations models. Readers interested in
including these functions into their population synthesis codes can
make use of the {\sc fortran} subroutine referred to in
Section~\ref{indexdef}.

Other results of this work are the following: i) We find a complex
behaviour of the Ca\,{\sc ii} strength as a function of the three main
atmospheric parameters ($T_{\rm eff}$, $\log g$ and [Fe/H]). For hot
and cold stars, effective temperature and luminosity class are the
main driving parameters, whereas, in the mid-temperature regime, all
three parameters play an important role (see
Fig.~\ref{fitCaTPaT}). ii) The residuals of the fitting functions
arise mainly from uncertainties in the input atmospheric
parameters. iii) We do not find any correlation between these
residuals and [Ca/Fe] abundance ratios.  iv) A comparison with the
fitting functions derived in the literature reveals striking
differences in the predicted gravity and metallicity dependences for
some regions of the parameter space (see Fig.~\ref{compfit}).

Basically, compared to previous fitting functions for the Ca\,{\sc ii}
triplet, the main advantage of our predictions is that they include
the whole range of effective temperatures. This is extremely important
since the Ca\,{\sc ii} behaviour in the temperature ranges explored by
the previous works can not be extrapolated at all to hot ($\theta \la
0.75$) or cold stars ($\theta \ga 1.35$), as it is readily seen in,
for instance, Fig.~\ref{triplot}. We should not forget that these cold
stars constitute a significant contribution to the integrated near-IR
spectra of most stellar populations (to old populations through the
role of cold dwarfs and giants, and to relatively young populations
through the expected contribution of AGB stars; see Paper~IV). This
must indeed be remarked since, if the fitting functions by the
previous works are extrapolated, stellar populations synthesis models
will derive unreliable integrated Ca\,{\sc ii} strengths which can not
be compared with observations.

In the case of synthesis models that make use of the theoretical
fitting functions of JCJ, the comparison of the predictions with the
observed Ca\,{\sc ii} triplet in galactic spectra is even more
uncertain since: i) JCJ predictions do not include the contamination
of Paschen lines, which affects the measurement of the observed index
if the contribution of warm stars is not negligible; and ii) JCJ
provides true equivalent widths, whereas line-strength indices in
systems like that of DTT are used for the observational data. Although
JCJ measure the strengths using the central bandpasses of DTT, these
measurements can not be compared to indices in this system, since the
local pseudo-continuum (traced between the sidebands) is far from
being a true continuum (especially for hot and cold stars; see
Paper~I).

As a final conclusion, we want to emphasize the importance of using an
empirical library with a wide range of accurate atmospheric parameters
and a sufficiently large number of calibrating stars for deriving
reliable empirical fitting functions.  Also, well defined index
definitions and objective fitting procedures are a critical factor.
To summarize, the functions presented in this paper alleviate these
problems and should lead to more reliable comparisons between models
and observations than in the past.  The first predictions of
population synthesis models using these results will be presented in
Paper~IV.

\section*{ACKNOWLEDGMENTS}

The authors are indebted to the referee G. Da Costa for useful
comments and suggestions. JC acknowledges the Comunidad de Madrid for
a Formaci\'on de Personal Investigador fellowship. AV acknowledges the
support of the PPARC rolling grant 'Extragalactic Astronomy and
Cosmology in Durham 1998-2002'. This research has made use of the
NASA's Astrophysics Data System Article Service. This work was
supported by the Spanish Programa Sectorial de Promoci\'on del
Conocimiento (PB96-610) and Programa Nacional de Astronom\'{\i}a y
Astrof\'{\i}sica (AYA2000-977) and, in part, by a British Council
grant within the British/Spanish Joint Research Programme (Acciones
Integradas).

\label{lastpage}

\end{document}